\documentclass[twocolumn, trackchanges]{aastex631} 

\usepackage[T1]{fontenc}

\usepackage[utf8]{inputenc}
\usepackage{enumitem}
\usepackage{float} 
\usepackage{gensymb}
\usepackage{amsmath}
\usepackage{textcomp}
\usepackage{verbatim}
\usepackage{tabularx}
\usepackage{placeins}
\usepackage{multirow}
\usepackage{graphicx}
\usepackage{placeins}
\usepackage{tgcursor}
\usepackage{enumitem}
\usepackage{chemfig}
\usepackage[version=4]{mhchem}

\usepackage{booktabs}

\newcommand{\Q}[1]{\textcolor{blue}{#1}}

\begin{document}

\title{Thermal Desorption Kinetics, Binding Energies, and Entrapment of Methyl Mercaptan Ices}
\shortauthors{Narayanan et al.}
\shorttitle{Thermal Desorption and Entrapment of CH$_3$SH Ices}


\author[0000-0002-0244-6650]{Suchitra Narayanan}
\altaffiliation{National Science Foundation Graduate Research Fellow}
\altaffiliation{P.E.O. Scholar}
\affiliation{Center for Astrophysics \textbar\ Harvard \& Smithsonian, 60 Garden St., Cambridge, MA 02138, USA}
\affiliation{Institute for Astronomy, University of Hawai`i at Mānoa, 2680 Woodlawn Dr., Honolulu, HI 96822, USA}

\author[0000-0001-6947-7411]{Elettra L. Piacentino}

\affiliation{Center for Astrophysics \textbar\ Harvard \& Smithsonian, 60 Garden St., Cambridge, MA 02138, USA}
\author[0000-0001-8798-1347]{Karin I. \"Oberg} 

\affiliation{Center for Astrophysics \textbar\ Harvard \& Smithsonian, 60 Garden St., Cambridge, MA 02138, USA}

\author[0000-0003-2761-4312]{Mahesh Rajappan}
\affiliation{Center for Astrophysics \textbar\ Harvard \& Smithsonian, 60 Garden St., Cambridge, MA 02138, USA}

\begin{abstract}
\noindent 
Organosulfur species are potential major carriers of sulfur in the interstellar medium, as well as interesting ingredients in prebiotic chemistry. The most fundamental question regarding these species is under which conditions they reside in the gas versus solid phase. {Here,} we characterize the thermal desorption kinetics, binding energies, and entrapment of the organosulfur methyl mercaptan (CH$_3$SH, or MeSH) in different ice environments, {comparing them with those of} methanol (CH$_3$OH, or MeOH) ices. {T}he {derived} multi-layer (pure MeSH$-$MeSH) and sub-monolayer (layered MeSH$-$H$_2$O) binding energies {are} surprisingly similar, corresponding to snow line locations where the disk midplane temperature is $\sim$\,105 K. {I}n both H$_2$O-dominated and more realistic H$_2$O:CO$_2$-dominated ices, 100\% of the MeSH {is} entrapped, almost exclusively desorb{ing} at the molecular volcano desorption peak, indicating that MeSH is retained {at} the water snow line if initially mixed {with} water ice during formation. {Additionally, the presence of MeSH in an ice mixture enhances the entrapment of CO$_2$ and MeOH (up to 100\%) until the onset of volcano desorption; without MeSH, both desorb at their respective pure desorption temperatures and also co-desorb with water.} Compared to MeOH, MeSH binds less well to water, explaining why MeSH escapes during water ice crystallization rather than co-desorbing with water. These results show the larger relative size of MeSH compared to MeOH significantly impacts its ability to bind to water and its entrapment efficiency{. Therefore,} molecular size plays a{n} important role in the adsorption and retention of S-bearing organics {and, in turn, other volatiles} in {ices}.\end{abstract}



\section{Introduction}
\label{sec:intro}




Sulfur (S), one of the elemental ingredients for life as we know it, is poorly understood compared to the other biologically-relevant elements in an interstellar context. One long-standing question is the so-called ``sulfur depletion problem'' where the gas-phase sulfur abundance is close to the Solar value in the diffuse regions of the interstellar medium but $\lesssim$\,$1$\% of the Solar value in the denser regions \citep{Tieftrunk1994, Ruffle1999, Howk2006, Goicoechea2006}. To reconcile this discrepancy, it is suggested that this ``missing'' sulfur is locked up in solid ices and/or refractory sulfur chains and minerals \citep{Smith1991, Laas2019}.  However, of the 25+ S-bearing molecules detected in space, only OCS has been confirmed in ices, along with tentative detections of SO$_2$ \citep{McGuire2022, Palumbo1995, Boogert1997, Palumbo1997, McClure2023}, and together these do not account for more than 5\% of the S budget \citep{Boogert1997, Palumbo1997}. {Recent works have shown that sulfur could be locked up in ammonium hydrosulphide (NH$_4$SH) salts, accounting for up to 18\% of the S \citep{Vitorino2024, Slavicinska2025}. O}rganic sulfur-bearing species (compounds containing H, C, S), have {also} been proposed as a possible sulfur reservoir in ices \citep{Laas2019}, but their chemistry and partitioning between the ice and gas is poorly constrained due to limited experimental work. However, a recent study investigating the formation of several S-bearing organics has shown that these molecules can act as effective sulfur sinks \citep{Santos2024}, demonstrating the need for more experiments to understand their\,behavior.

In this paper, we focus on the simplest complex\footnote{Following the convention set in \citet{Herbst2009}, we define a molecule with six or more atoms to be complex.} organosulfur, methyl mercaptan (also called methanethiol or CH$_3$SH). CH$_3$SH is of particular interest because it is thought to form like its oxygen-bearing counterpart, methanol (CH$_3$OH), which is relatively abundant in ices \citep{Boogert2015, McGuire2022, McClure2023}. Though unobserved in ices, CH$_3$SH has been detected in the gas phase across many environments ranging from the Sgr B2 and OMC-1 molecular cloud complexes \citep{Linke1979, Turner1991}, to a variety of cloud cores, L1544, B1, and G327.3-0.6  \citep{Vastel2018, Cernicharo2012, Gibb2000a}, to protostars, W33A and IRAS 16293-2422 \citep{Gibb2000a, Majumdar2016}. Additionally, CH$_3$SH has been found in comet 67P/Churyumov-Gerasimenko, further motivating understanding its behavior in ices \citep{Calmonte2016}. CH$_3$SH is also of astrobiological importance as it is both a precursor to two S-bearing amino acids \citep{vanTrump1972, Heinen1996} and a potential biosignature in exoplanets \citep{Pilcher2003, Schwieterman2018}.

Characterizing the CH$_3$SH reservoir during planet formation requires a detailed understanding of its formation and destruction pathways, as well as the disk conditions under which it is present in the gas or ice, i.e. its snow line location(s). The latter is directly governed by binding energies and entrapment. The binding energy ($E_b$) measures how strongly a molecule binds to a particular surface, and entrapment refers to when a  volatile molecule is `trapped' within less volatile ice matrices. Existing constraints on CH$_3$SH binding energies comes from CH$_3$SH sublimation off {of} a gold surface \citep{Liu2002} and {quantum chemical} calculations \citep{Wakelam2017, Perrero2022}. There is currently no experimental study of CH$_3$SH entrapment within ice matrices. Since there can be a \mbox{$\sim$\,100–2000 K} difference between experimentally-derived and theoretically-predicted binding energies \citep{Wakelam2017, Piacentino2022}, and because water ices have been shown to effectively entrap several volatile species \citep{Bar-Nun1985, Collings2003, Collings2004}, investigating these properties experimentally for key molecules highly desirable.

To address how small organosulfurs behave in ices, this paper presents results from thermal desorption experiments that quantify the binding energies and entrapment efficiencies of CH$_3$SH ices—both of which are vital to astrochemical models and interpretation of observations. We also present every CH$_3$SH experiment with its CH$_3$OH counterpart to obtain a better mechanistic understanding of what factors contribute to CH$_3$SH's behavior and in which ways the O and S organic reservoirs may differ. The rest of the paper is organized as follows: \S\ref{sec:expts} describes the experimental methods; \S\ref{sec:results} presents the derived binding energies and entrapment efficiencies; \S\ref{sec:discussion} discusses the physical chemistry and astrophysical implications of our results; and finally, \S\ref{sec:conclusions} summarizes our main conclusions.

\section{Experimental Details}
\label{sec:expts}

\subsection{Experimental Setup}

All experiments presented in this paper were conducted on the ultra-high vacuum (UHV) chamber SPACE–KITTEN\footnote{Surface Processing Apparatus for Chemical Experimentation–Kinetics of Ice Transformation in Thermal ENvironments} which is described in detail in \cite{Simon2023}. The UHV chamber is pumped down to a base pressure of $\sim$\,$4\times10^{-9}$ Torr at room temperature ($\sim$\,298 K). In the center of the chamber is a 2\,mm thick cesium iodide (CsI) substrate that is transparent to infrared (IR) radiation and can be cooled down to 14 K via a closed-cycle helium cryostat. The substrate temperature is controlled and monitored by a LakeShore Model 335 temperature controller that is calibrated to an absolute accuracy of $\sim$\,2 K with a relative uncertainty of $\pm$\,0.1 K. To grow ices, gases are introduced into the chamber via a gasline with a base pressure of $<$\,$5\times10^{-4}$ Torr, which in turn leads to a doser positioned  $\sim$\,1 inch from the substrate. A Bruker Vertex 70 Fourier transform infrared spectrometer (FTIR)  operated in the transmission mode is used to measure the abundance of IR-active species present in the ice. In order to obtain the composition of the gas-phase species, a Pfeiffer QMA 220 PrismaPlus quadrupole mass spectrometer (QMS)  is used. The mass fragments chosen for monitoring via QMS are selected by inspecting the mass spectrum for each molecule in the NIST Chemistry WebBook and choosing the most abundant peak \citep{NIST}.

\begin{deluxetable*}{llccccc}[ht!]
\tablecaption{IR band strengths used to calculate the column density of experimental ices. \label{tab:bandstrengths}}
\tablewidth{\textwidth}
\tablehead{
    \colhead{Molecule} & \colhead{Chemical Formula} & \colhead{Mode} & \colhead{Position} & \colhead{Band Strength$^{*}$ (A$^\prime$)} & Temperature  & \colhead{Reference}  \\[-2mm]
    \nocolhead{dummy} & \nocolhead{dummy}   & \nocolhead{dummy} & \colhead{[cm$^{-1}$]} & \colhead{[cm molecule$^{-1}$]} & \colhead{[K]}  & \colhead{}
    }
\startdata 
Methyl Mercaptan$^\star$& $^{13}$CH$_3$SH &  S$-$H stretch &  2535 & $5.41 \times 10^{-18}$ & 17  & 1\\
Methanol$^\star$ & $^{13}$CH$_3$OH & C$-$O stretch & 1028$^{\dagger}$   & $ 1.62 \times 10^{-17}$ & 10  & 2\\
Water & H$_2$O & O$-$H stretch & 3280 & $2.20 \times 10^{-16}$ & 14  & 3,\,4$^{\ddagger}$\\
Carbon Dioxide & CO$_2$ & C$-$O stretch  & 2343 &  $1.10 \times 10^{-16}$ & 14  & 3,\,4$^{\ddagger}$\\
\hline
\hline
\enddata
\tablecomments{$^{*}$We assume a 20\% error for all band strengths.\\$^{\star}$The band strengths for $^{13}$C- are unavailable so the respective $^{12}$C- values are used.\\$^{\dagger}$For $^{13}$CH$_3$OH, the peak position is shifted to 1000 cm$^{-1}$.
\\$^{\ddagger}$We use the density-corrected value from \citet{Bouilloud2015} which is based on \citet{Gerakines1995}. }
\tablerefs{1. \citet{Hudson2016}; 2. \citet{Hudson2024}{;} 3. \citet{Gerakines1995}; 4. \citet{Bouilloud2015}{.}}
\end{deluxetable*}

\subsection{Chemical Reagents and Preparation}
\label{sec:chemicals}

The gaseous chemicals used in this work are $^{13}$CH$_3$SH (MilliporeSigma; 99\% isotopic purity,  97\% chemical purity) and $^{12}$CO$_2$ (MilliporeSigma; 99.9\% purity), which were used directly from the lecture bottles with no further purification. We use $^{13}$C-methyl mercaptan out of necessity; at the time of the onset of these experiments the normal isotopologue was not available. We assume that the derived properties are valid for both $^{13}$C and the normal isotopologue, since the mass difference is only a few percent and previous experiments on the effect of $^{12}$C and $^{13}$C isotopologues on binding energies have found only small differences that are well within our derived uncertainties \citep{Smith2021}.  The liquid chemicals used are $^{13}$CH$_3$OH (MilliporeSigma; 99\% isotopic purity, 99\% chemical purity), which we select to match the methyl mercaptan, and deionized water (H$_2$O). Both were transferred into evacuated flasks and further purified through at least three freeze-pump-thaw cycles using liquid nitrogen. For ease of readability, hereafter we refer to $^{13}$CH$_3$SH as MeSH {and}  $^{13}$CH$_3$OH as MeOH. We use the term MeXH when referring to either MeSH or MeOH experiments, where X represents S or O.

\subsection{Experimental Procedures}
\label{sec:procedures}

We performed three types of thermal desorption experiments: \textit{(1)} multi-layer (single-component MeXH ices), \textit{(2)} sub-monolayer (layered ices where MeXH is deposited on compact amorphous water), and \textit{(3)} mixed ices (binary, MeXH:H$_2$O, and ternary, MeXH:CO$_2$:H$_2$O {and MeSH:MeOH:H$_2$O}).  In brief, an experiment starts with first cooling the CsI substrate down to the desired deposition temperature. For the mixed ices, the vapors were introduced into the gasline and allowed to settle for $\sim$\,5 minutes. Both the pure (i.e., single-component) and the mixed ices were deposited directly at 14 K. During dosing, the ice and gas species are constantly monitored using the FTIR and QMS, respectively, until the desired ice coverage (in monolayers, ML, where we follow convention and set 1 ML equal to 10$^{15}$ molecules cm$^{-2}$) is reached. For details on how the ice coverage is estimated for different regimes, see \S\ref{sec:icecoverage_calc}.

For all layered experiments, the base (i.e. molecule in contact with the substrate) is compact H$_2$O. {Compared to porous ices, compact ice substrates minimize entrapment of surface volatiles, enabling relatively clean binding energy measurements. However, our discussion of sub-monolayer MeOH$-$H$_2$O experiments (see \S\ref{sec:subML}) shows that differentiating between entrapment and binding energies for molecules with pure desorption temperatures close to water can become challenging. While compact water ices are commonly used as laboratory models for interstellar ice grains, there are some potential differences. Theoretical models suggest that ices should be largely compact \citep{Garrod2013}. However, recent observations suggest that ices may be somewhat more porous than previously thought \citep[see e.g.,][]{McClure2023, Noble2024}, though additional data is needed to confirm. Given these uncertainties, we opted for a compact H$_2$O substrate for the sub-monolayer to reduce the effects of entrapment when studying binding interactions with water ice surfaces.} 

To ensure a compact amorphous ice structure, H$_2$O was deposited at 100 K {at a normal incidence}. Following this step, the substrate is cooled back down to 14\,K and MeXH is slowly deposited at a controlled rate of $<$\,1 ML per minute. Once dosing is complete, the ice sample is subjected to temperature programmed desorption (TPD), where the substrate is heated at a constant rate of 2~K~min$^{-1}$ until all species are fully desorbed off the substrate ($\sim$\,200–250 K). The QMS measurements are taken for a particular mass-to-charge ratio $m/z$ as a function of temperature, and the resulting TPD curves serve as the foundation for this work. Thus, the substrate {is rotated to face the QMS} {to optimize QMS measurements, prioritizing} gas-phase monitoring over {obtaining} IR spectra during {heating}. The post-dosing (pre-heating) FTIR spectra used to calculate the initial ice coverage and the TPD curves make up the experimental data products.\footnote{All data products are available on Zenodo at \texttt{{\url{https://doi.org/10.5281/zenodo.13827075}}} \citep{NarayananZenodo}.}

\subsection{Ice Coverage Calculation}\label{sec:icecoverage_calc}

Using the spectrum obtained from the FTIR, the column density of IR-active molecules in the ice is determined using the following equation:

\begin{equation}\label{eqn:columndens}
    N_x =\frac{\int_\text{band} \tau_x(\Tilde{\nu}) d\Tilde{\nu}}{A^\prime_x} , 
\end{equation}

where $N_x$ is the ice column density of a specific molecule $x$ in molecules cm$^{-2}$, $\int\tau_x(\Tilde{\nu}) d\Tilde{\nu}$ is the integrated optical depth over the wavenumber $(\Tilde{\nu})$ range of the IR band in absorbance units, and $A^\prime_x$ is the band strength in cm\;molecule$^{-1}$ \citep{Hudgins1993}. The band strengths of all molecules studied in this work are listed in Table \ref{tab:bandstrengths}. Since there are no literature values for $^{13}$CH$_3$SH, we use the available band strengths for $^{12}$CH$_3$SH.  We also adopt the $^{12}$CH$_3$OH band strength for the C-O stretch to calculate the $^{13}$CH$_3$OH column densities, since this mode was unambiguously methanol in the layered and mixed experiments. For the limited number of molecules where the properties of both the $^{12}$C- and $^{13}$C-isotopologues have been investigated, the band strengths of features that are impacted directly by the isotopic substitution ({i.e.}, C–O stretches) exhibit variations of no more than 5–10\% \citep[e.g., C–O stretch for CO and CO$_2$ in][]{Bouilloud2015}. {A recent study suggests that these differences may be higher: \citet{Gerakines2023} find a 56\% difference for $^{12}$C- and $^{13}$CO, but only 5\% for  C-$^{16}$O  and $^{17}$O and 26\% for C-$^{16}$O and C$^{18}$O, which is 2 amu higher. However, even if these values are confirmed by other studies, we expect the differences for the heavier MeSH to be smaller, especially when considering the S–H stretch, which should be minimally affected by the presence of $^{12}$C- or $^{13}$C. This assumption is tested and validated using quantum chemical calculations, where we model the geometries, binding energies and band strengths of each isotopologue (see computational details in Appendix \ref{app:g16}). We also check the sensitivity of our results to this error (see \S\ref{sec:nu_TST} for details) and do not find substantial changes to the result. We therefore adopt a band strength uncertainty of 20\% as our default.} Representative pure and mixed spectra for $^{13}$CH$_3$SH can be found in Appendix\,\ref{app:spectra}.

Unlike for the multi-layer and mixed ice experiments, we cannot confirm the sub-monolayer coverage from the initial post-dose IR spectrum because of insufficient starting material to be able detect the main MeXH feature. We find the detection limit to be $\sim$\,2 ML for MeSH (see Appendix \ref{app:spectra}). In fact, we confirmed our experiments were in the sub-monolayer regime using the absence of the MeXH feature after dosing (pre-analysis). To determine the sub-monolayer ice coverages, we created a calibration curve using the pure desorption experiments and fitting for a calibration constant that relates the QMS response (at the desired $m/z$) to the IR-derived column density (calculated using Eq. \ref{eqn:columndens}). As ice thicknesses increase, complexities arise in the QMS response; therefore, we derive the calibration constant using only the three thinnest MeSH experiments, as these are most relevant for sub-monolayer analyses. The integrated QMS TPD signal from the sub-monolayer experiments is then scaled by the calibration constant to recover the initial ice coverage. All calibration curves to determine the scaling factor are found in Appendix \ref{app:subML_calc}.

\section{Results}
\label{sec:results}
\begin{figure}
    \centering
     \includegraphics[width = \columnwidth]{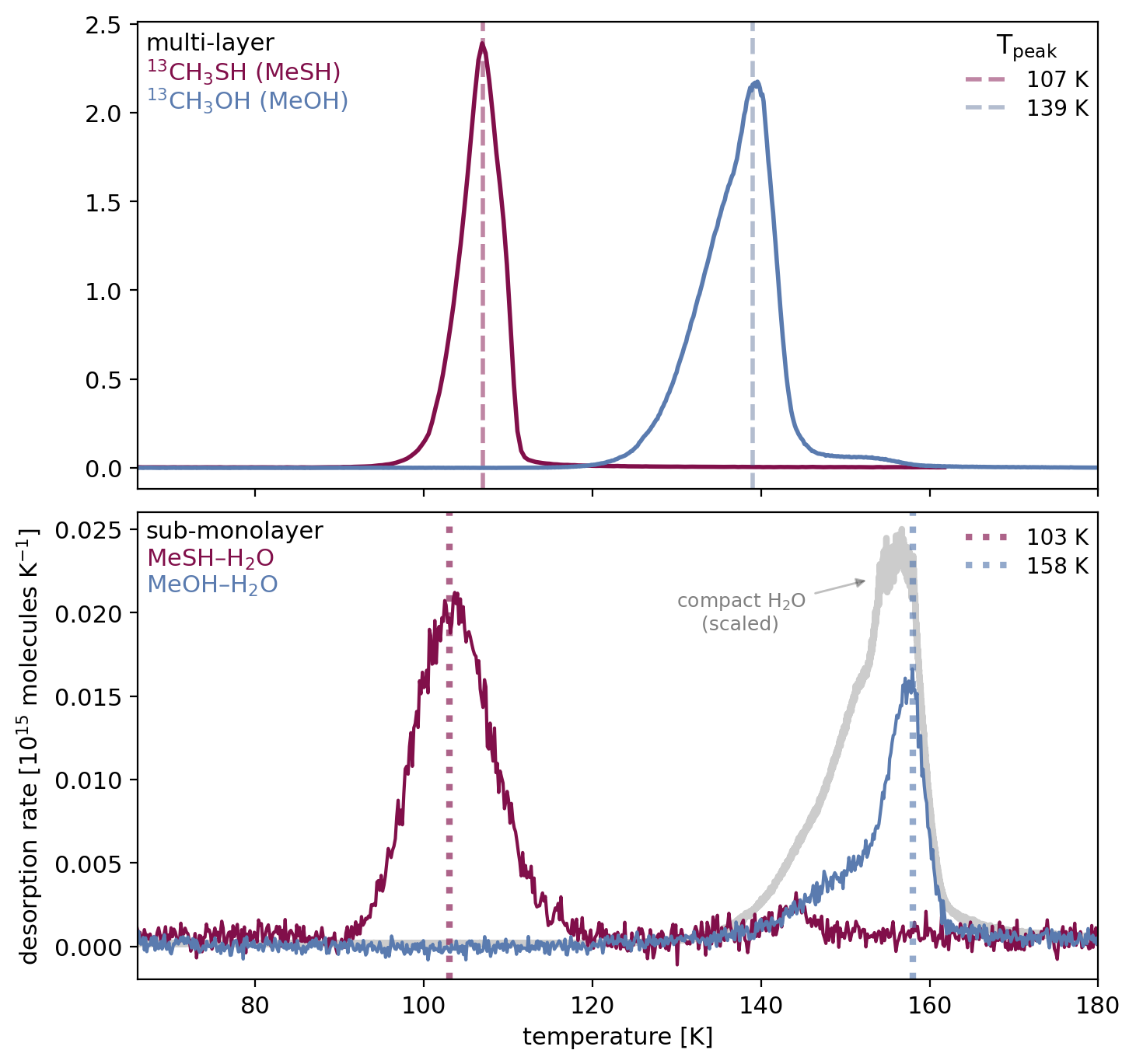}
    \caption{Temperature programmed desorption (TPD) curves of MeSH (magenta) and MeOH (blue) ices in the multi-layer (top panel) and sub-monolayer (bottom panel) regimes, corresponding to Expts. 3, 8, 10, and 15 in Table \ref{tab:thermal_expt_summ}. The dashed and dotted lines represent the temperature at which a particular profile peaks (or T$_\text{peak}$). For the sub-monolayer experiments, a compact H$_2$O TPD curve (normalized to the peak value and then scaled by $\frac{1}{40}$) is shown in gray for easy reference.} \label{fig:comparison_tpds}
\end{figure}

Figure \ref{fig:comparison_tpds} shows representative TPD curves in the multi-layer and sub-monolayer regimes of MeSH and MeOH. In both cases, MeSH desorbs before MeOH, indicating that MeSH is more volatile and its binding energy is always lower than that of MeOH. In the sub-monolayer regime, MeSH does not appear to bind more strongly to water than to itself, while MeOH does. Additionally, there is only minimal entrapment of MeSH when deposited on top of water, indicating MeSH barely makes it into the pores that exist at the surface of a compact amorphous water ice, while in the case of MeOH there are signs of significant entrapment. {As is elaborated in \S\ref{sec:subML}, it is difficult to determine whether the sub-monolayer TPD profile of MeOH is indicative of MeOH binding to the compact water substrate, as a consequence of its ability to form strong hydrogen bonds, or if it is due to entrapment.}

In the following sections we investigate these characteristics quantitatively using a series of multi-layer (\S\ref{sec:mult}), sub-monolayer (\S\ref{sec:subML}), and entrapment experiments. The summaries of all of the thermal desorption and entrapment experiments used to extract the binding energies and entrapment efficiencies are presented in Table \ref{tab:thermal_expt_summ} and \ref{tab:entrap_expt_summ}, respectively. The recommended binding energies are found in Table \ref{tab:rec_BEs}.

\begin{figure}
    \centering
     \includegraphics[width = \columnwidth]{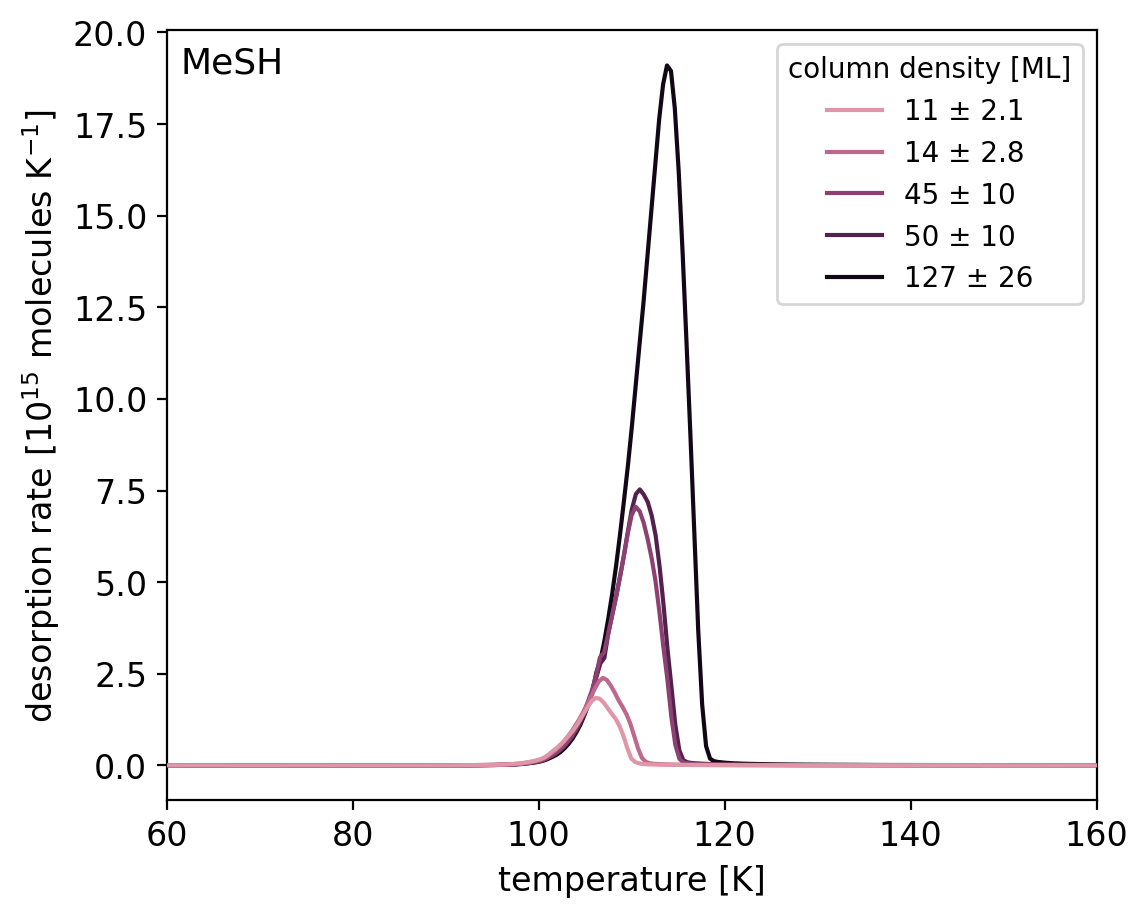}
    \caption{TPD curves of the pure MeSH experiments (Expts. 2–6 in Table \ref{tab:thermal_expt_summ}) used to obtain multi-layer binding energies. Note that as the column density increases, so does the T$_\text{peak}$, while the leading edges remain coincident.} \label{fig:full_MeSH_multilayer_tpds}
\end{figure}

\subsection{Multi-layer Binding Energies}
\label{sec:mult}

The TPDs of all multi-layer experiments used to extract MeSH$-$MeSH binding energies are shown in Figure \ref{fig:full_MeSH_multilayer_tpds}. While the desorption peak temperature (T$_\text{peak}$) increases as a function of ice coverage, the leading edges of all curves are coincident, as expected if the desorption is {of} zeroth order. Given that all experiments are done in the UHV regime where readsorption is negligible, we can derive the binding energy by fitting the leading edge of TPDs to the Polanyi-Wigner equation \citep{Polanyi1925},  
\begin{equation}\label{eqn:PW}
    -\frac{d\theta}{dT} = \frac{\theta^n\nu}{\beta}\exp\bigg[-\frac{E_{b}}{T}\bigg],
\end{equation}

where $\theta$ is the ice coverage in ML, $T$ is the ice temperature in K, $d\theta/dT$ is the desorption rate in ML K$^{-1}$, $n$ is the desorption order, $\nu$ is pre-exponential factor in ML$^{(1-n)}$ s$^{-1}$, $\beta$ is the heating rate in  K min$^{-1}$, and $E_{b}$ is the binding energy (also sometimes denoted as $E_\text{des}$ for desorption energy) in K. For pure multi-layer ices, we fit Eq. \ref{eqn:PW} using zeroth order kinetics ($n$\,=\,0) to obtain $E_{b}$. The pre-exponential factor, $\nu$, typically referred to as the attempt frequency, quantifies the number of attempts per second a molecule makes to escape the ice matrix. We fit for $\nu$ using three methods described in detail below: direct fitting of experimental TPD curves ($\nu_{\mathrm{expt}}$), the harmonic oscillator approximation ($\nu_{\mathrm{harm}}$), and the transition state theory (TST) model ($\nu_{\mathrm{TST}}$). The resulting zeroth-order Polanyi-Wigner fits to determine $\nu$ and multi-layer (MeXH–MeXH) $E_b$ is presented in Figures \ref{fig:MeSH_combined_PW_fits} for MeSH. All corresponding figures for MeOH can be found in Appendix \ref{app:MeOH_supp}.

\begin{figure}
    \centering
     \includegraphics[width = \columnwidth]{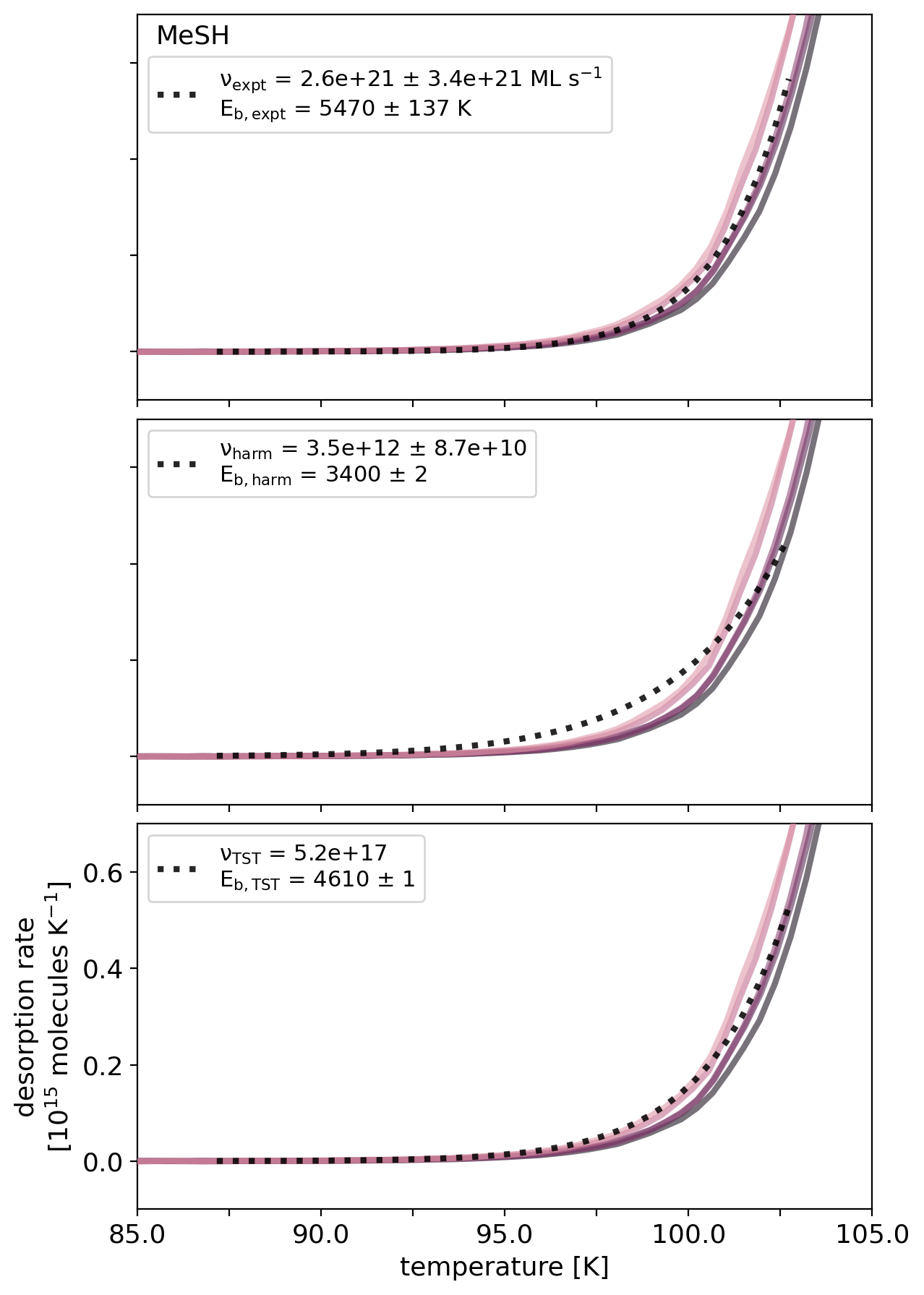}
    \caption{Zoomed-in view of the leading edges of the experimental TPD curves for pure MeSH ices shown in Figure \ref{fig:full_MeSH_multilayer_tpds}. The dotted black line shows the zeroth-order Polanyi-Wigner fit over the temperature region where there is overlap between all curves. The top, middle, and bottom panels show the fits using the experimental, harmonic, and TST approximations to estimate $\nu$ and $E_b$ (described in detail in \S\ref{sec:mult}), respectively. The standard errors of the fit are also shown for each of the parameters.} \label{fig:MeSH_combined_PW_fits}
\end{figure}

\begin{deluxetable}{ccc}
\tabletypesize{\footnotesize}
\tablecaption{Summary of Desorption Experiments}
\label{tab:thermal_expt_summ}
\tablehead{\colhead{Expt.} & \colhead{Ice Type} &
\colhead{Column Density$^a$ [ML]}}
\startdata
~\,1$^*$ & pure$^b$  MeSH  & 2.3 $\pm$ 0.7 \\
2  &  & 11 $\pm$ 2.1  \\
3  &  & 14 $\pm$ 2.8  \\
4  &  & 45 $\pm$ 10  \\
5  &  & 50 $\pm$ 10 \\
6  &  & 127 $\pm$ 26 \\
\hline
7 & pure  MeOH  &  4.7 $\pm$ 1.0 \\
8  &  & 21 $\pm$ 5  \\
9  & & 36 $\pm$ 8 \\
\hline
\hline
10 & layered  MeSH on H$_2$O & 0.25 $\pm$ 0.06 \\
11 &  & 0.45 $\pm$ 0.10   \\
12 &  & 0.78 $\pm$ 0.17 \\ 
\hline
13 &  layered MeOH on H$_2$O & 0.12 $\pm$ 0.02 \\
14 & & 0.16 $\pm$ 0.03  \\
15 & &  0.34 $\pm$ 0.07  \\
16 & &  0.71 $\pm$ 0.14  \\
\hline
\hline 
\enddata
\tablecomments{$^{*}$Calibration only experiment (see Appendix \ref{app:subML_calc}).\\$^{a}$The column density is for the MeXH species. For layered experiments, the column density of compact H$_2$O is $\approx$30–40 ML, which is sufficiently thick to ensure the sub-monolayer ice interacts solely with the compact H$_2$O substrate.\\$^{b}$Pure refers to single-component mixtures.\vspace{-0.4in}}
\end{deluxetable}

\subsubsection{Empirical Estimation}\label{sec:nu_expt}

To obtain $\nu_{\mathrm{expt}}$ empirically, we directly fit the TPD curves to Eq. \ref{eqn:PW} using a non-linear least-squares minimization method  \citep[i.e.\,{\fontfamily{lmtt}\selectfont{lmfit}},][]{lmfit} that simultaneously constrains $E_b$ and $\nu$ following \citet{Piacentino2024}.  Since both values are highly degenerate, we fit all pure desorption profiles together, under the assumption that both parameters are independent of ice coverage; this assumption is further discussed in \S\ref{sec:comp} and Figure \ref{fig:MeSH_BETSTafoML}.

The resulting empirical fit (top panel of Figure \ref{fig:MeSH_combined_PW_fits})  shows that the error in $\nu_{\mathrm{expt}}$ exceeds itself, while the standard error on $E_{b,\mathrm{expt}}$ appears better constrained. In reality the error in $E_{b,\mathrm{expt}}$ is considerably higher, since it changes substantially with small shifts in fitting region. Although overall the formal errors for the respective MeOH fits are smaller, the $E_{b,\,\mathrm{expt}}$ is predicted to be lower for MeOH than MeSH, in contrast with what we qualitatively expect based on the pure MeXH desorption temperatures. Thus, we conclude that our experiments are not sufficient to break degeneracy between $\nu_\mathrm{expt}$ and $E_{b,\,\mathrm{expt}}$. Since it is clear that $\nu_\mathrm{expt}$ and $E_{b,\,\mathrm{expt}}$ are poorly constrained via this method, we do not derive formal errors as they are not representative of the true uncertainties of the extracted values.


\subsubsection{Harmonic Oscillator Approximation}\label{sec:nu_harm}

The harmonic oscillator approximation of $\nu$ \citep[see][]{Hasegawa1992}, which has been used in many previous studies and is commonly implemented in astrochemical codes, is described by the equation

\begin{equation}\label{eqn:nu_harm}
    \nu_\mathrm{harm} = \sqrt{\frac{2N_sE_{b,\,\mathrm{harm}}}{\pi^2\mu m_\text{H}}},
\end{equation}

where $N_s$ is the binding site density (fixed at 10$^{15}$ cm$^{-2}$), and $\mu m_\text{H}$ is the mean molecular weight. Using Eq. \ref{eqn:nu_harm}, we iteratively solve for the binding energy $E_{b,\,\mathrm{harm}}$ using the  {\fontfamily{lmtt}\selectfont{lmfit}} minimizer.

The harmonic  approximation performs the worst, visually, of the three methods, as seen by the deviation of the best-fit curve from the leading edges in the middle panel of Figure \ref{fig:MeSH_combined_PW_fits}. The poor fit indicates that the harmonic approximation, which assumes the molecule is a point mass, is not an appropriate model for MeSH. This method also does not appear to be a good approximation for MeOH, though the divergence is smaller than in the case of MeSH. Nevertheless, while the MeXH species may not be well-described as a point mass, we include the harmonic approximation due to its well-established use in the literature and implementation in astrochemical models  
\citep[e.g., {\fontfamily{lmtt}\selectfont{NAUTILUS}}, {\fontfamily{lmtt}\selectfont{ALCHEMIC}};][]{Ruaud2016, Semenov2010}. This allows for comparison with previous studies and provides a basis for evaluating its capabilities against the other methods. We do not derive formal errors from this fit as these would underestimate the problems with this approximation. 

\begin{deluxetable}{lcccc}
\tablecaption{Recommended TST-derived binding energies and pre-exponential factors.}
\label{tab:rec_BEs}	
\tabletypesize{\footnotesize}
\tablehead{\colhead{}  &
\colhead{$n$}  & \colhead{$E_{b,\,\mathrm{TST}}$ [K]} & 
\colhead{$\nu_\text{TST}$$^a$} & \colhead{T$_\text{peak}$$^b$ [K]}} 
\startdata
MeSH$-$MeSH & 0 & 4610 $\pm$ 110 & 5.2$\substack{+2.8 \\ -1.0}$\,$\times$\,10$^{17}$ &   $106\substack{+14 \\ -6}$  \\
MeSH$-$H$_2$O & 1&   4640 $\pm$ 170 &4.9$\substack{+0.6\\-0.9}$\,$\times$\,10$^{17}$  & 104 $\pm$ 5  \\
\hline 
MeOH$-$MeOH &  0&  5750\,$\pm$\,80 & $3.4\substack{+1.5 \\ -0.9}$\,$\times$\,10$^{17}$ & $131\substack{+14 \\ -11}$ \\
\hline
\hline 
\enddata
\tablecomments{For why we are unable to derive  a binding energy for \mbox{MeOH$-$H$_2$O}, please refer to Appendix \ref{app:MeOH_submonolayer}.\\$^{a}$Units are ML s$^{-1}$ for zeroth-order ($n$\,=\,0) and s$^{-1}$ for first-order ($n$\,=\,1) desorption.
\\$^{b}$We set the recommended T$_\text{peak}$ to be the one corresponding to the thinnest multi-layer ice used for $E_b$ analysis, while the $\pm$ values represent the range of peak temperatures spanning the experimental ice coverages, rather than strict errors.\vspace{-0.3in}}
\end{deluxetable}
\vspace{-0.13in}

\subsubsection{{TST Approximation}}\label{sec:nu_TST}
A more accurate method for estimating $\nu$ for bigger molecules is the TST model described in detail in \cite{Minissale2022}, which accounts for the  partition function of the desorbing species. This approach only considers the translational and rotational degrees of freedom, since the desorption temperatures for the molecules studied in this work are insufficient to require consideration of the excited vibrational and/or electronic states. The translational ($q^{\ddagger}_{tr, \text{2D}}$) and rotational ($q^{\ddagger}_{rot,\text{3D}}$) partition functions are calculated using equations 

\begin{equation}\label{eqn:qtr}
q^{\ddagger}_{tr, \text{2D}}=A \bigg[\frac{h}{\sqrt[]{2 \,\pi\, m \,k_B \,T_\text{peak}}} \bigg]^{-2}
\end{equation}

and 
\begin{equation}\label{eqn:qrot}
    q^{\ddagger}_{rot, \text{3D}}=\frac{\sqrt[]{\pi}}{\sigma \,h^3} (8\, \pi^2 \,k_B\, T_\text{peak})^{3/2} \sqrt[]{I_x\, I_y \,I_z}, 
\end{equation}

where A is the surface area per adsorbed molecule (fixed to 10$^{-19}$ m$^2$), $h$ is the Planck constant, $m$ is the mass of the particle, $k_B$ is the Boltzmann constant, and $T_\text{peak}$ is as previously defined (temperature at which the TPD curve peaks). The symmetry factor, $\sigma$, and principal moments of inertia (I$_x$, I$_y$, and I$_z$) were determined computationally, the details of which are in  Appendix \ref{app:g16}. We can then derive $\nu_\text{TST}$ using the equation 

\begin{equation}
    \nu_\text{TST}=\frac{k_B T_\text{peak}}{h}q^{\ddagger}_{tr, \text{2D}} \; q^{\ddagger}_{rot, \text{3D}}.
\end{equation}

\begin{figure}[ht!]
\centering
\includegraphics[width=\columnwidth ]{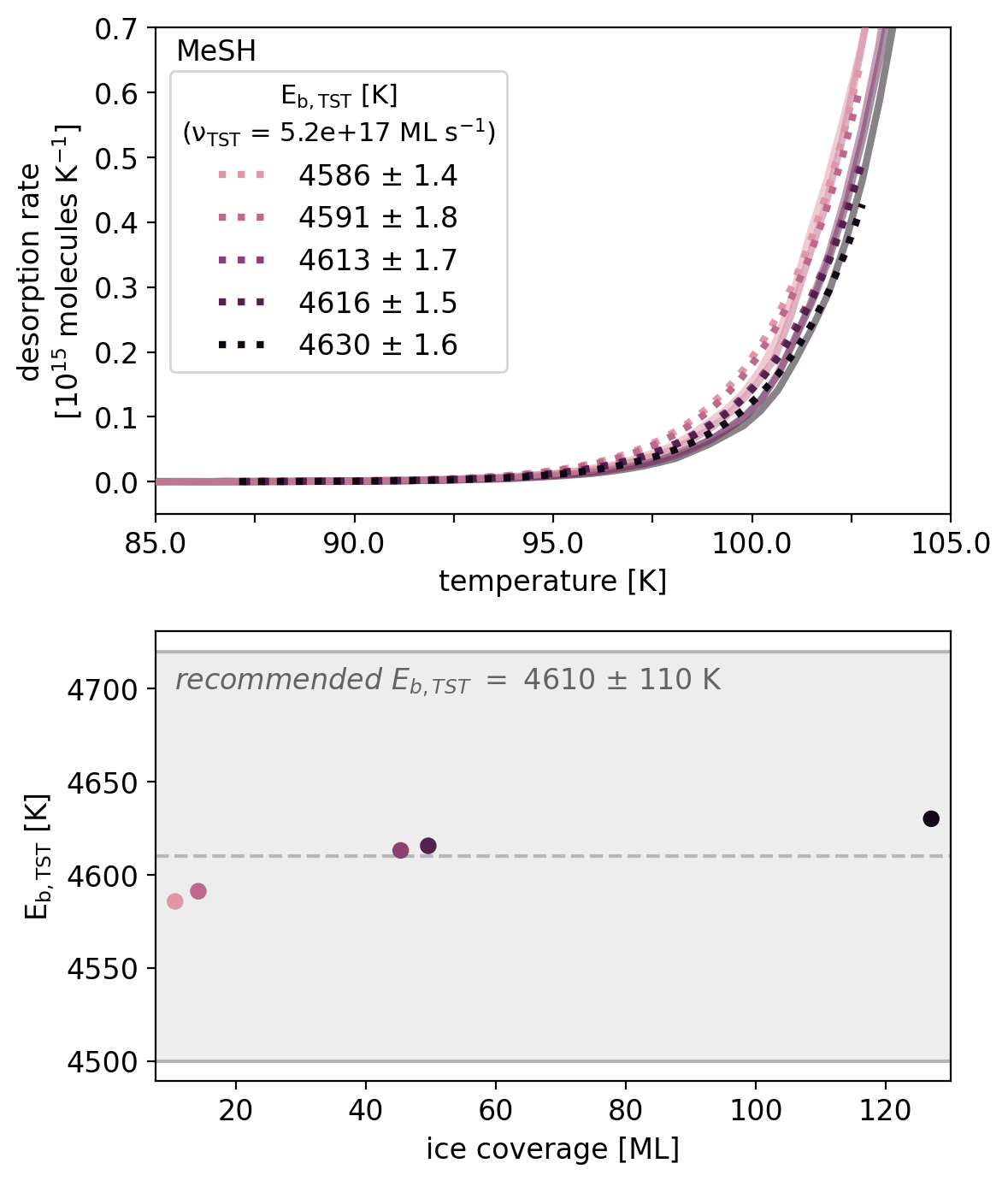}
\caption {\textit{Top:} Similar to the bottom panel of Figure \ref{fig:MeSH_combined_PW_fits}, except each multi-layer TPD curve is individually fitted for. The individual zeroth-order Polanyi-Wigner fits are depicted by the corresponding colored dotted line and use the same $\nu_\mathrm{TST}$ (determined by using the T$_\mathrm{peak}$ of the thinnest ice being fit for). The extracted $E_{b,\,\mathrm{TST}}$ values are shown with only their standard errors of fit. \textit{Bottom:} Individually fitted $E_{b,\,\mathrm{TST}}$ using the same $\nu_\mathrm{TST}$ as a function of ice coverage. There is a weak dependence, but note that the overall range of the extracted $E_{b,\,\mathrm{TST}}$ values is $\sim$\,50 K, well-within our recommended uncertainty of $\pm$\,110 K as presented in Table \ref{tab:rec_BEs}. The recommended value is depicted as a dashed gray line, with the associated uncertainties represented by the shaded gray region for clearer visualization.} \label{fig:MeSH_BETSTafoML}
\end{figure}

Note that $\nu_\text{TST}$ is independent of the binding energy and is entirely theoretical with the exception of needing $T_\text{peak}$ from the experimental data.

The bottom panel in Figure \ref{fig:MeSH_combined_PW_fits} shows the TST fit, which both aligns with the leading edges well and results in the lowest standard error.  To estimate a more formal error, we consider several sources of uncertainty that affect the TST-derived attempt frequency and binding energy: the value(s) chosen for T$_\mathrm{peak}$, uncertainties in band strengths and therefore ice coverages, the temperature range over which the fit is performed, uncertainties in substrate temperatures, and the formal error from the fits. We also consider possible differences in binding energies between $^{12}$C and $^{13}$C isotopologues.

For $\nu_\mathrm{TST}$, the only source of error arises from the way T$_\mathrm{peak}$ is defined. We selected T$_\mathrm{peak}$ based on the thinnest ices used in the multi-layer analysis (corresponding to Expts. 2 and 7 in Table \ref{tab:thermal_expt_summ}). However, since T$_\mathrm{peak}$ varies as a function of ice coverage, we considered temperatures across the fitting range, resulting in $\nu_\mathrm{TST}$ uncertainties of 20-50\% as listed in Table \ref{tab:rec_BEs}. {While the TST method was developed for the sub-monolayer regimes \citep{Ligterink2023}, our analysis suggests that this approximation remains reasonable in the multi-layer regime because the $\nu_\mathrm{TST}$ uncertainties, which take into account changes in T$_\mathrm{peak}$ as a function of coverage, contribute minimally to the overall derived $E_{b,\,\mathrm{TST}}$ uncertainties (see below).}

We systematically tested each source of uncertainty that affects the $E_{b,\,\mathrm{TST}}$ individually to check if one source dominated over the others.  The uncertainty on coverage was tested by varying the value of the ice coverage by 20\%. For the fitting region, we kept the lower temperature limit the same (as we found no dependency of the resulting fits on the lower bound), and varied the upper temperature limit to range when the temperature began to curve upward, indicating onset of desorption, up to the T$_\text{peak}$ of the thinnest multi-layer ice. In both cases, the resulting $E_{b,\,\mathrm{TST}}$ varied by at most 10\,K. The formal fit errors were only at most $\sim$\,2 K, marking the smallest contribution. {We also individually fit the multi-layer ices using their respective $\nu_\mathrm{TST}$, calculated from their T$_\text{peak}$, and found variation of $\sim$\,10–20 K.} While we could not directly test the isotopologue contribution ourselves, we estimate the difference to only be $\sim$\,10–15 K, based on previous work showing a binding energy difference of 15 K between $^{12}$CO and $^{13}$CO, where the 1 amu mass difference has a greater impact due to the smaller size of the molecule \citep{Smith2021}. {We do computationally verify that the isotopologue does not change the binding energy; there is no difference in either the energies or optimized binding geometries for either isotopologue (see details in  Appendix \ref{app:g16}).}  

We found that the dominant source of uncertainty is the absolute temperature uncertainty, which we tested by varying the temperature data by $\pm2$\,K. This contributed to $\sim$\,80–100\,K difference in  $E_{b,\,\mathrm{TST}}$, significantly more than the other sources of uncertainty. Thus, these are the uncertainties presented in Table \ref{tab:rec_BEs}. {To also check whether the $^{12}$C- vs. $^{13}$CO band strength difference found in \citet{Gerakines2023} affects our results, we varied the ice coverage by 56\% and found that the resulting $E_{b,\,\mathrm{TST}}$ fits were within the errors presented in  Table \ref{tab:rec_BEs}.} Additionally, to explore the dependence of $E_{b,\,\mathrm{TST}}$ on the ice coverage, we fit each multi-layer curve individually using the same $\nu_\mathrm{TST}$, which is shown in Figure \ref{fig:MeSH_BETSTafoML}, showing a slight dependence but well within our recommended values.

\subsubsection{Comparison and Recommendation}\label{sec:comp}

In summary, the empirical and harmonic approximations methods do not work well for MeSH because of the degeneracy between $E_b$ and $\nu$, and a{n oversimplified} physical model, respectively. By contrast, the TST method performed well for both MeSH and MeOH and we therefore recommend that these values (found in Table \ref{tab:rec_BEs}) are used. Using the TST method, we find that the MeSH$-$MeSH $E_b$ is lower than MeOH$-$MeOH $E_b$ by 1140 K.

\begin{deluxetable*}{clccccccccc}
\tablecaption{Summary of Entrapment Experiments}
\label{tab:entrap_expt_summ}
\tablewidth{\textwidth} 
\tablehead{\colhead{Expt.} &
    \colhead{Ice Composition} & \multicolumn{3}{c}{Column Density [ML]}  & \colhead{Total} &
    \colhead{Ratio} &  \multicolumn{2}{c}{MeXH$_\text{trap}$ [\%]} & \multicolumn{2}{c}{{CO$_2$$_\text{\,trap}$ [\%]}} \\   
      \cmidrule{3-5}\cmidrule(l{2pt}r{2pt}){8-9} \cmidrule(l{2pt}r{2pt}){10-11}
    \colhead{} & 
    \colhead{}  & 
    \colhead{{MeXH}}  &
    \colhead{{CO$_2$}} & 
    \colhead{{H$_2$O}} & [ML] & \colhead{} &\colhead{{$^{a}$H$_2$O$_\text{vol}$}} & \colhead{{$^{b}$H$_2$O$_\text{tot}$}}&\colhead{{$^{a}$H$_2$O$_\text{vol}$}} & \colhead{{$^{b}$H$_2$O$_\text{tot}$}} }
\startdata
{17} & MeSH:H$_2$O  & 1.2 $\pm$ 0.3 &  -- &  74 $\pm$ 15 & 75 & 1:{62} & 100 & 100 & {--} & {--}  \\
{18} & & 8.5 $\pm$ 1.7 &  -- &  87 $\pm$ 18 & 96 & 1:{10} & 100 & 100 & {--} & {--}\\
\hline
{19} & MeSH:CO$_2$:H$_2$O   &   1.9 $\pm$ 0.5 & 2.5 $\pm$ 0.5 & 9.3 $\pm$ 1.9 & 14 & 1:{1}:{5} & 100 & 100 & {41} & {55} \\
{20} & & 6.1 $\pm$ 1.2 & 14 $\pm$ 2.8 & 72 $\pm$ 15  & 92 & 1:2:12 & 100 & 100 & {59} & {76} \\
{~21$^\dagger$} & & 9.0 $\pm$ 1.8 & 18 $\pm$ 4 &  75 $\pm$ 16  & 103 & 1:{2}:{8}& 100 & 100 & {--} & {--}  \\
{22} & & 9.4 $\pm$ 1.9 & 35 $\pm$ 7 &  54 $\pm$ 11  & 98 & 1:{4}:{6} & 52 &  52 & {16} & {22}  \\
\hline 
\hline
{23} & MeOH:H$_2$O    & 8.9 $\pm$ 1.8&  -- &  87 $\pm$ 18 & 96  & 1:{10} &  {39} &  {48} & {--} & {--} \\
\hline
{24} & MeOH:CO$_2$:H$_2$O  &  6.0 $\pm$ 1.2 & 12 $\pm$ 2.5 & 63 $\pm$ 13  & 82 & 1:{2}:{11} & 43 &77& {8} & {43} \\
\hline\hline
& & {} & {} & {} & & & \multicolumn{2}{c}{{MeSH$_\text{trap}$ [\%]}} & \multicolumn{2}{c}{{MeOH$_\text{trap}$ [\%]}} \\   
& & {MeSH} & {MeOH} & {H$_2$O} & & &\colhead{{$^{a}$H$_2$O$_\text{vol}$}} & \colhead{{$^{b}$H$_2$O$_\text{tot}$}}&\colhead{{$^{a}$H$_2$O$_\text{vol}$}} & \colhead{{$^{b}$H$_2$O$_\text{tot}$}}  \\   
  \cmidrule{3-5} \cmidrule(l{2pt}r{2pt}){8-9} \cmidrule(l{2pt}r{2pt}){10-11}
{25} & {MeSH:MeOH:H$_2$O}  &  {1.5 $\pm$ 0.4 } & {6.0 $\pm$ 1.2}  & {52 $\pm$ 10}  & {60} & {1:{4}:{35}} & {98} & {98}  & {66} & {99}\\
{26} &  &  {5.0 $\pm$ 1.0} & {40 $\pm$ 8}  & {166 $\pm$ 33}  & {211} & {1:{8}:{33}} & {100} & {100} & {90} & {100}\\
{27} &  &  {10 $\pm$ 2.0} & {14 $\pm$ 2.8}  & {62 $\pm$ 12}  & {86} & {1:{1.4}:{6}} & {93} & {98} & {87} & {96}\\
{28} &  &  {12 $\pm$ 2.5} & {22 $\pm$ 4} & {145 $\pm$ 29}  & {178} & {1:{2}:{12}} & {92} & {93} & {90} & {98}\\
\hline
\hline
\enddata
\tablecomments{All mixed ices were deposited at 14 K. We assume entrapment efficiency errors of $\sim$\,5\% based on previous replicate entrapment experiments \citep{Simon2023}, {except in the case of 100\% entrapment, in which there are no quantifiable errors}.\\{$^{a}$}\,{Amount of volatile entrapped in the volcano desorption peak}.\\{$^{b}$}\, 
{Total amount of volatile entrapped (including both volcano desorption and co-desorption with H$_2$O).}\\{$^\dagger$For this experiment, we were unable to obtain the CO$_2$ entrapment efficiencies due to incorrect data collection.}\\}
\end{deluxetable*}

\vspace{-0.23in}

\subsection{Sub-monolayer Binding Energies}\label{sec:subML}

\begin{figure}
    \centering
     \includegraphics[width = \columnwidth]{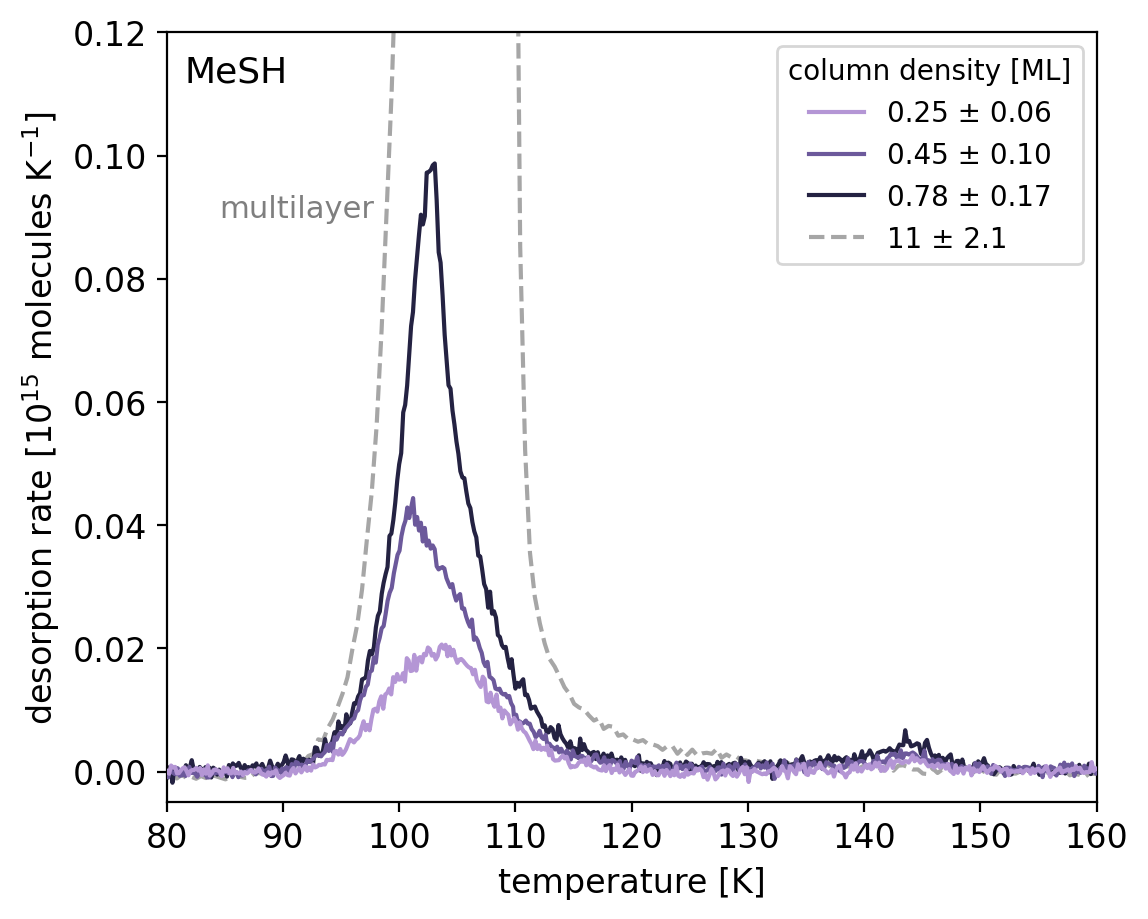}
    \caption{TPD curves of the sub-monolayer MeSH on compact H$_2$O experiments (Expts. 9–11 in Table \ref{tab:thermal_expt_summ}).  Each curve exhibits a slight bump at around 145\,K associated with the crystallization temperature of water, corresponding to slight entrapment of MeSH within the surface pores of {the} compact H$_2$O layer. This represents an upper limit of MeSH entrapment within the compact H$_2$O surface since {t}he height of the bump does not increase significantly as a function of ice thicknesses. The multi-layer TPD curve corresponding to Expt. 2 is shown as a dashed gray line for easy comparison.}\label{fig:full_MeSH_subML_tpds}
\end{figure}

\begin{figure}
    \centering
     \includegraphics[width = \columnwidth]{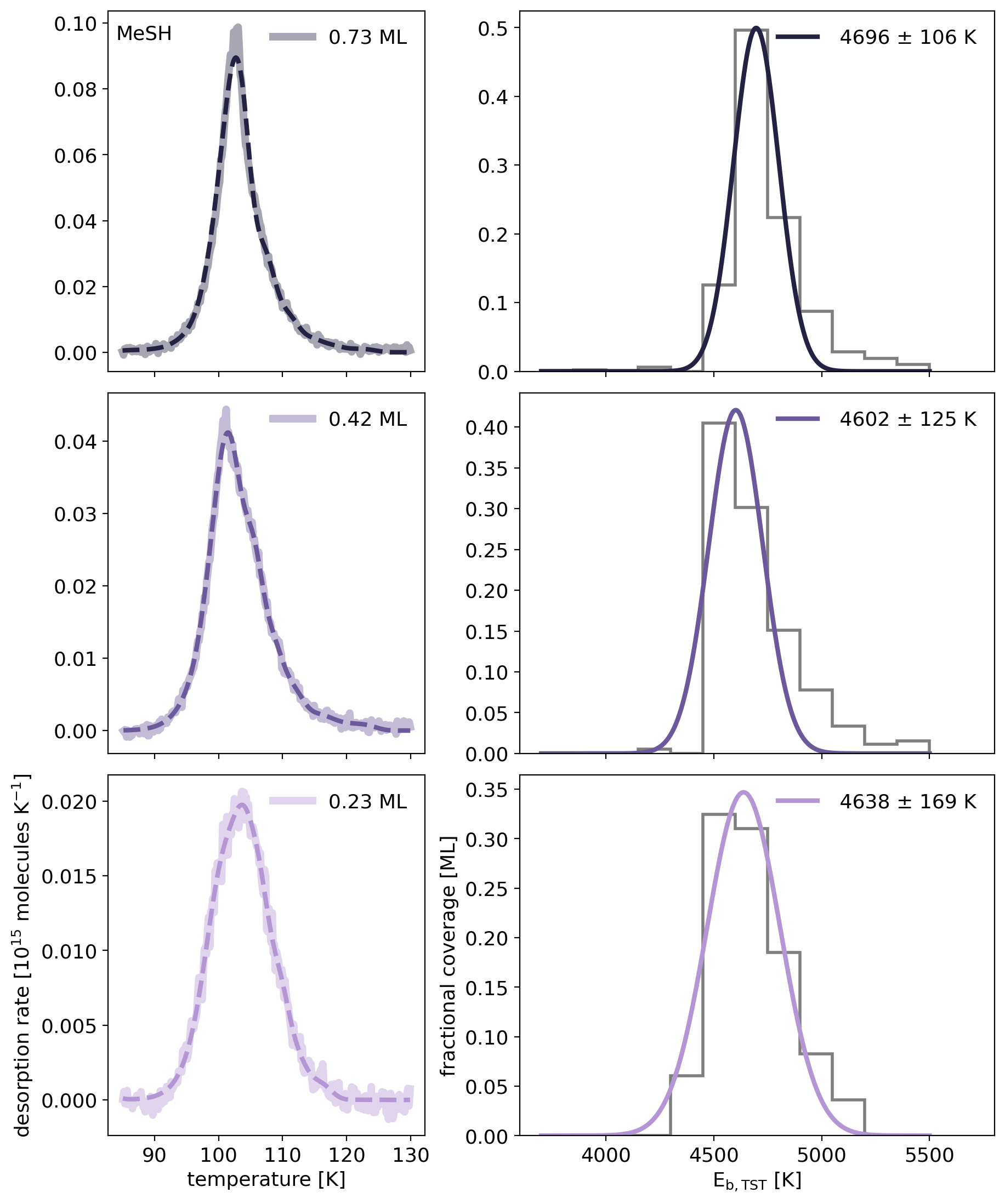}
    \caption{\textit{Left:}  Individual sub-monolayer (MeSH on compact H$_2$O) TPD curves corresponding to Expts. 10–12 in Table \ref{tab:thermal_expt_summ} from top to bottom. The column density presented is the effective ice coverage. The dashed lines are the corresponding first-order Polanyi-Wigner fits for a distribution of binding energies. \textit{Right:} Corresponding binding energy distributions represented as histograms of the fractional coverages with a smoothed fit (solid line) assuming a Gaussian distribution.}\label{fig:MeSH_subML_PW_fits} 
\end{figure}

Figure \ref{fig:full_MeSH_subML_tpds} shows the TPDs of all sub-monolayer experiments, used to extract MeSH$-$H$_2$O binding energies. As the ice coverage decreases, the peak temperature barely shifts, but the peak shape becomes more symmetric, indicative of a more complete transition from multi-layer to true sub-monolayer desorption kinetics.

In the sub-monolayer regime, we fit Eq. \ref{eqn:PW} using first-order kinetics ($n$\,=\,1). We solve the resulting ordinary differential equation (ODE), $d\theta/dT$, using {\fontfamily{lmtt}\selectfont{odeint}} from {\fontfamily{lmtt}\selectfont{SciPy}} \citep{2020SciPy-NMeth}, and fit to the entire desorption curve, simultaneously solving for $\theta_0$ and $E_{b,\,\mathrm{TST}}$, where the fitted coverage is expected to be smaller than the calibrated ice coverages due to the possibility of some MeSH molecules being present on top of one another in island-like structures. We calculate the T$_\text{peak}$ by fitting a Gaussian to the desorption curve and use the full-width half maximum (FWHM) to estimate the uncertainty of this value, which results in a calculated $\nu_\text{TST}$ of 4.9$\substack{+0.6\\-0.9}$\,$\times$10$^{17}$ s$^{-1}$ which is similar to the multi-layer $\nu_\text{TST}$, within uncertainties.

We solved the ODE initially assuming a single binding energy for MeSH$-$H$_2$O, which did not fit the data well, indicative of a non-uniform surface that has a distribution of binding sites. Instead, we fit the sub-monolayer curve using a distribution of binding energies by sampling a range of  $E_{b,\,\mathrm{TST}}$ from 3800–5000 K, and modeling the binding energies as a linear combination of first-order desorption kinetics \citep[see, e.g.,][]{Noble2012, Fayolle2016, Behmard2019}. These results are shown in Figure \ref{fig:MeSH_subML_PW_fits}. All experiments result in similar binding energies, but due to the more symmetric profile of the thinnest sub-monolayer MeSH experiment and a better distribution fit, we recommend the MeSH$-$H$_2$O binding energy to be 4640\,$\pm$\,170\,K, where the ``error'' denotes the width of the binding energy distribution, which is larger than the sources of uncertainty described in \S\ref{sec:nu_TST}.

There is no substantial impact of MeSH and H$_2$O interactions on the binding energy, which is in sharp contrast with MeOH (see Appendix \ref{app:MeOH_submonolayer}) where the sub-monolayer TPD curves becomes coincident at the trailing edge (at 160 K) and align with the compact amorphous water curve. The curves are also asymmetric and appear to reflect multiple  distinct desorption regimes even in the thinnest experiments. Due to these complex asymmetries and aligned trailing edges that coincide with water co-desorption, we are unable to derive MeOH$-$H$_2$O, as it is ambiguous whether the desorption features result from a true sub-monolayer MeOH interaction with the compact water substrate or co-desorption (i.e. entrapment) with water, though it appears the latter case is what dominates in the MeOH$-$H$_2$O TPD curves.

\subsection{Entrapment in Mixed Ices}
\label{sec:entrapment}
\begin{figure*}[ht!]
    \centering
     \includegraphics[width = \textwidth]{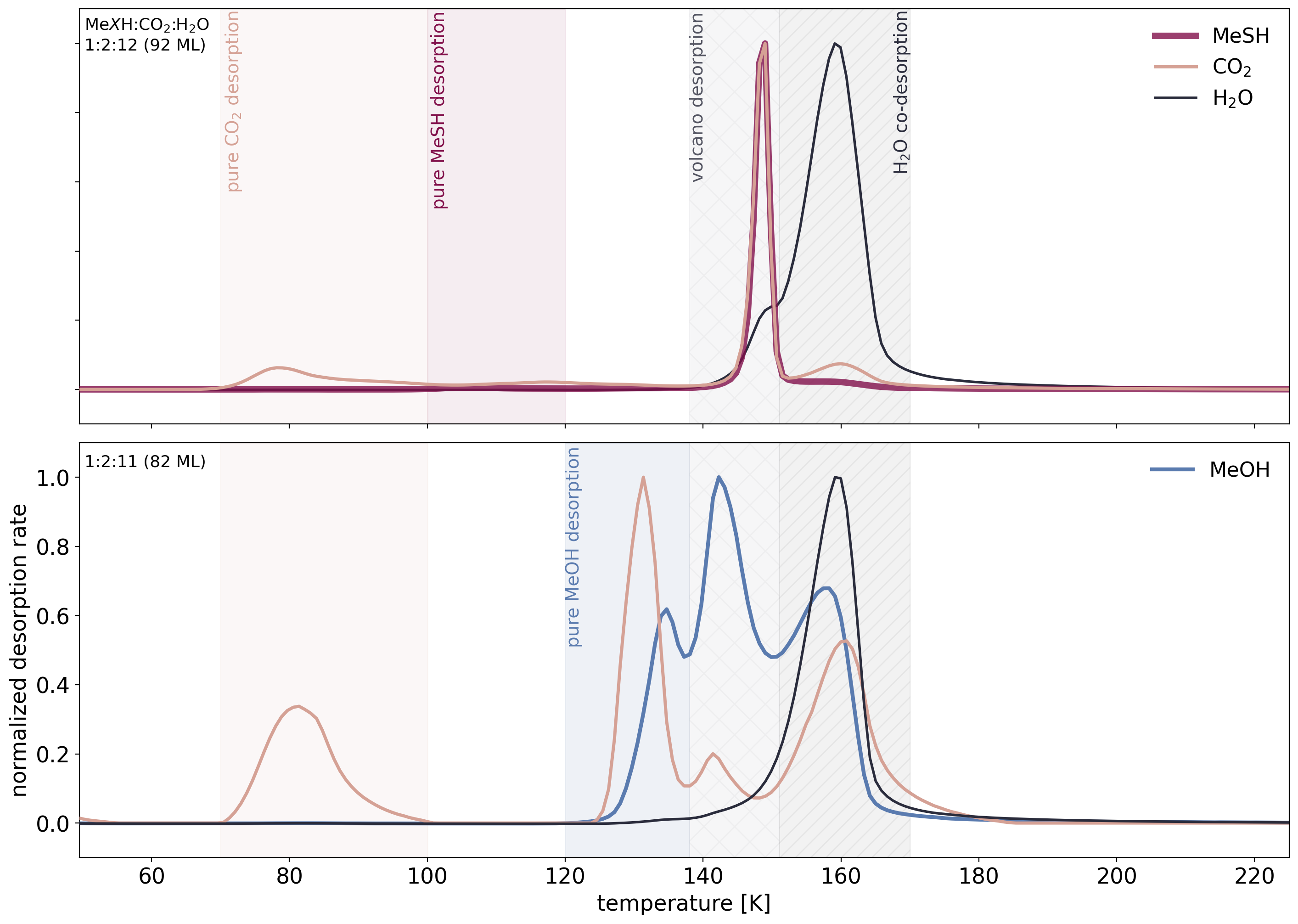}
    \caption{Normalized TPD curves of fiducial ternary (MeXH:CO$_2$:H$_2$O) entrapment experiments of MeSH (top panel) and MeOH (bottom panel), corresponding to Expts. 2{0} and 2{4} in Table \ref{tab:entrap_expt_summ}. The temperature ranges corresponding to different desorption zones of particular molecule is shaded, and these are the bounds are used to quantify the entrapment efficiencies {presented in Table \ref{tab:entrap_expt_summ}}. } \label{fig:norm_entrap_comparison} 
\end{figure*}

The entrapment experiments were performed in \textit{(i)}~binary water-rich matrices (MeXH:H$_2$O),  \textit{(ii)}~more astrophysically-realistic ternary mixtures (MeXH:CO$_2$:H$_2$O) based on observations of interstellar clouds and protostellar envelopes  \citep{Allamandola1999, Boogert2015}, {and \textit{(iii)}~ternary mixtures of MeSH:MeOH:H$_2$O}. The fiducial cases {for \textit{(i)} and \textit{(ii)}} are $\sim$100 ML experiments with composition ratios of $\sim$\,1:10 and $\sim$\,1:2:12 for the binary and ternary mixtures, respectively. The details of these and the additional experiments varying ratios and thicknesses are listed in Table \ref{tab:entrap_expt_summ}. Figure \ref{fig:norm_entrap_comparison} shows the two fiducial ternary mixtures with all curves normalized to 1 to better visualize the different desorption peaks for MeSH and MeOH. The regions corresponding to different desorption temperatures of species are shaded; a molecule is considered to be entrapped if it desorbs after its normal (i.e. pure) desorption temperature. The `molecular volcano' desorption refers to the temperature at which H$_2$O transitions from crystalline to amorphous \citep{Smith1997}. This restructuring allows for abrupt desorption of underlying entrapped/volatile molecule(s). In the fiducial ternary case, we find that a 100\% of the MeSH is entrapped and comes off nearly completely at the volcano peak, whereas MeOH is only 77\% entrapped, and comes off at all three (pure, volcano and H$_2$O co-desorption) temperatures. The shaded regions are the bounds used to calculate entrapment efficiencies in the volcano region and in total (volcano + H$_2$O {co-}desorption) {which can be found in Table \ref{tab:entrap_expt_summ}.}

\begin{figure}
    \centering
     \includegraphics[width =\columnwidth
     ]{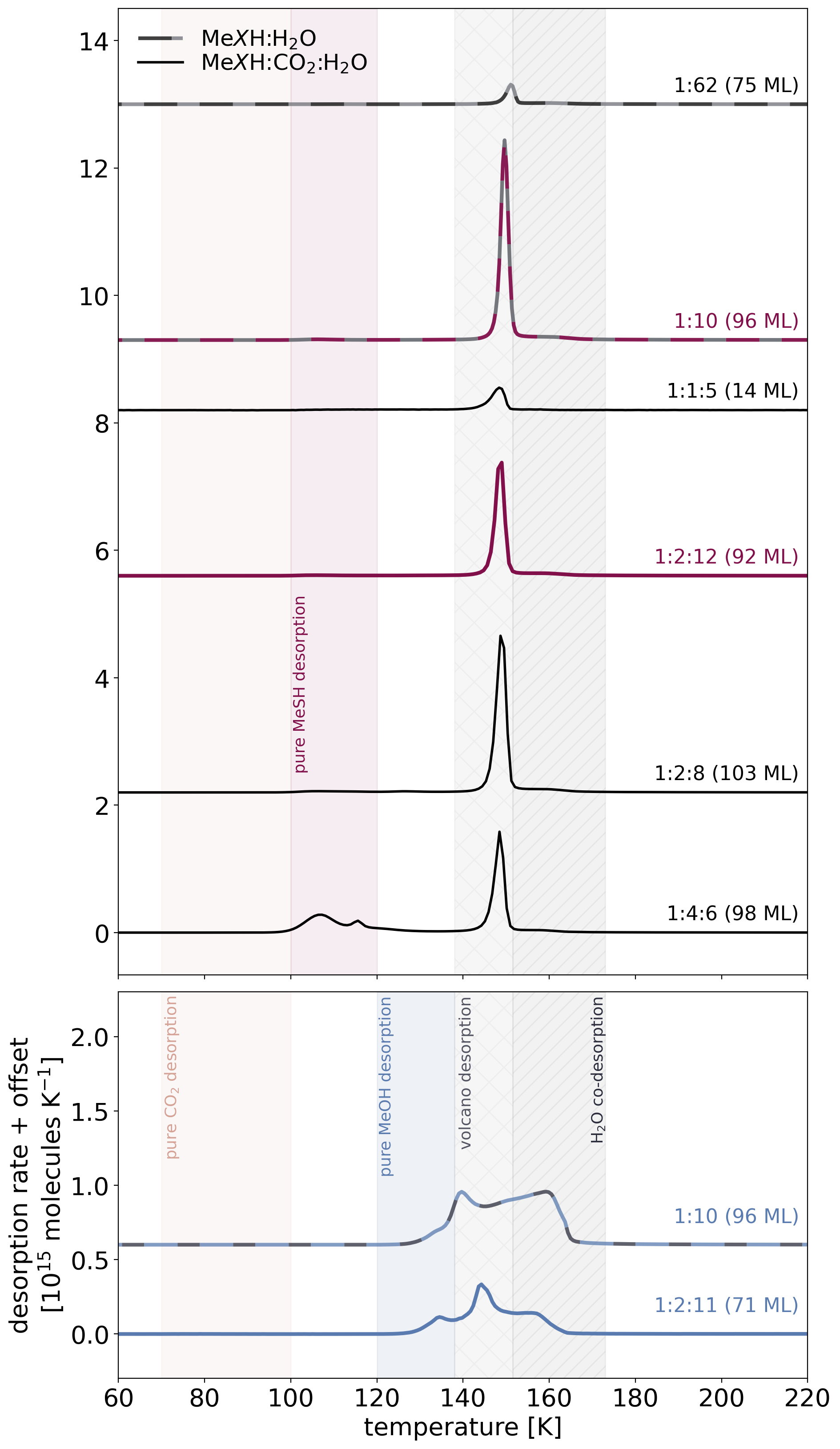}    \caption{ Unscaled {MeXH} TPD curves of all MeSH entrapment experiments (top) and fiducial MeOH experiments (bottom). The shaded regions are the same as in Figure \ref{fig:norm_entrap_comparison}. The curves (and labels) that are colored with their respective molecule color (magenta for MeSH and blue for MeOH) represent the fiducial binary (if dashed) and ternary (if solid) for easy comparison. \textit{Top:}\,\,TPD curves of all MeSH entrapment experiments ordered from {very dilute} binary mixtures to volatile-rich ternary mixtures corresponding to Expts. 17–2{2} (in order from top to bottom) in Table \ref{tab:entrap_expt_summ}. \textit{Bottom:}\,\,Fiducial binary and ternary {MeOH} entrapment experiments corresponding to Expts. 2{3} and 2{4}.} \label{fig:full_entrap_comparison} 
\end{figure}

In Figure \ref{fig:full_entrap_comparison} we show all of the entrapment TPD curves for MeSH (top panel) and the fiducial MeOH (bottom panel), unscaled. In the upper panel, from top to bottom the MeSH mixture conditions are moving from very dilute binary mixtures (Expt. 17 in Table \ref{tab:entrap_expt_summ}) to volatile-rich ternary mixtures (Expt. 2{2}). For MeOH, we show {the} fiducial binary and ternary experiments (Expt. 2{3} and 2{4}).  Consistent to what was shown in Figure \ref{fig:norm_entrap_comparison} for the fiducial experiments, MeOH generally desorbs at all three temperatures (corresponding to pure, volcano, and H$_2$O desorption), while MeSH is a 100\% entrapped and desorbs in the volcano region in all water-dominated ices. Even in the most volatile-rich experiment, we find that MeSH desorption kinetics is dominated by volcano desorption. Compared to MeOH, MeSH is curiously both better entrapped, in that MeSH desorption from ice mixtures is generally negligible, and less entrapped, in that co-desorption with water is much less important. {Additionally, the presence of MeSH results in more CO$_2$ sublimating at the volcano desorption when comparing entrapment efficiencies of CO$_2$ in the fiducial MeSH ternary experiment to its MeOH analog; when in a MeSH:H$_2$O matrix, 59\% of CO$_2$ comes off at the volcano peak, whereas only 8\% comes off in the analogous MeOH:H$_2$O mixture. Furthermore, when comparing total entrapment in water, 76\% of CO$_2$ is entrapped in the MeSH ternary matrix, whereas only 43\% is retained in the MeOH ternary.} Based on previous replicate entrapment experiments, we expect entrapment efficiencies to fluctuate by about 5\% {due to experimental errors} \citep{Simon2023}. 

\begin{figure}
    \centering
     \includegraphics[width =\columnwidth
     ]{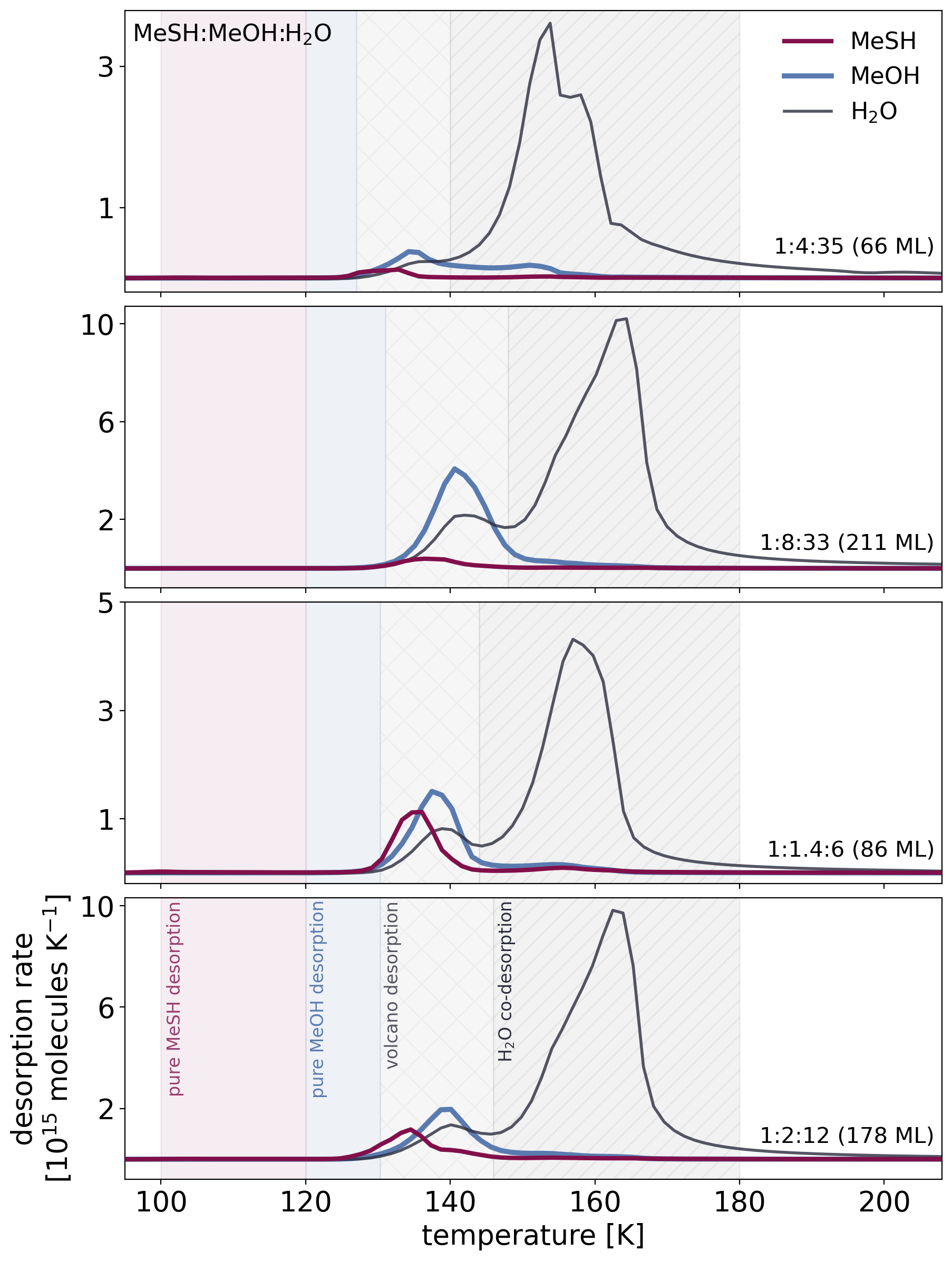}  \caption{{Unscaled TPD curves of all components of the ternary MeSH:MeOH:H$_2$O entrapment experiments. The shaded regions for MeOH, volcano, and H$_2$O co-desorption vary for each experiment as MeOH affects the H$_2$O crystallization kinetics, evidenced by the differences in H$_2$O TPD curve shapes and temperature shifts where restructuring occurs. The panels are ordered from top to bottom corresponding to Expts. 25–28 in Table \ref{tab:entrap_expt_summ}.} }\label{fig:MeXH_entraps} 
\end{figure}

{Finally, to explore how MeSH and MeOH desorb in the presence of each other in a water matrix, we ran ternary experiments of MeSH:MeOH:H$_2$O,  shown in Figure \ref{fig:MeXH_entraps}, which  correspond to Expts. 25–28 in Table \ref{tab:entrap_expt_summ}. In these experiments, MeOH is nearly 100\% entrapped regardless of composition, ratio and thickness, with MeOH desorption nearly exclusively at the volcano desorption peak. This is in stark contrast to the MeOH desorption and entrapment behavior in the ice analogs without MeSH. MeSH continues to desorb almost exclusively at the volcano desorption peak, though compared to the experiments without MeOH, MeSH desorption begins slightly earlier, preceding the onset of both MeOH and amorphous water desorption.} 

Of note is that the inclusion of MeOH appears to affect the water ice crystallization kinetics in the MeOH-rich {and organic-rich} entrapment experiments. {In the former}, there is no clear water desorption peak at the crystallization temperature {and in the latter the temperature at which restructuring occurs shifts}. This effect has been reported in previous studies \citep{Burke2015, Kruczkiewicz2024}. In contrast, MeSH does not appear to affect the water crystallization or amorphous desorption kinetics in any of the experiments, indicative of a weaker interaction between MeSH and water.

\section{Discussion}\label{sec:discussion}

\subsection{MeSH vs. MeOH Desorption Kinetics and Behavior}

For the pure, multi-layer case, MeSH is found to have a lower binding energy to itself than MeOH. This can be understood from the ability of MeOH to form {strong} hydrogen bonds, {especially relative to} MeSH {which exhibits very weak H-bonding} \citep{Kosztolanyi2003}. This is consistent with calculations of dimer interactions (see  Appendix \ref{app:dimer}). Notably, our analysis and another recent study using the TST Method  \citep{Kruczkiewicz2024} both yield significantly higher multi-layer MeOH binding energies than previous work utilizing the harmonic approximation ($E_{b,\,\text{harm}} = 4235 \pm 15$\,K) from \citet{Sandford1993}, highlighting that implementing the TST method can shift $E_b$ by $>$\, 1000\,K. However, the $E_{b}$ value presented in \citet{Kruczkiewicz2024} is \mbox{$\sim$\,220 K} lower compared to ours, probably due to a combination of choices with respect to peak temperature and fitting region, as well as the total number experiments studied. Note that our estimate is based on a larger experimental dataset, and therefore we recommend the multi-layer MeOH $E_b$ value in Table \ref{tab:rec_BEs} be used in future astrochemical studies.

In the sub-monolayer regime, we found that the binding energies of MeSH to itself and MeSH to water are \mbox{indistinguishable}, whereas MeOH binds more strongly to water, though we could not quantify the magnitude of the shift. To understand computationally what the difference between the multi-layer and sub-monolayer binding energies are, we performed dimer calculations of MeXH to a single H$_2$O  molecule {(see Appendix \ref{app:dimer} for more details)}, which has been shown to generally match experiments \citep{Piacentino2022}. For the MeSH dimers, the binding energy for MeSH$-$H$_2$O is higher than MeSH$-$MeSH, while for the MeOH dimers, MeOH$-$H$_2$O is slightly lower than MeOH$-$MeOH, which is opposite to the experimental trends. Possible reasons for this discrepancy are that dimers are not a good enough description of this system where long-range interactions might be important and/or that MeSH absorbed on a water surface cannot take advantage of the relatively large dimer interaction due to topological constraints. In other words, we speculate that MeSH may be too large to effectively bind to multiple water molecules on the surface. Topological constraints may also explain the higher MeOH binding energy {in} the sub-monolayer case if {this increase is due to} MeOH inserting itself into and strongly binding to nanopores present on the surface. {This increase in binding energy could also be due to the cooperative effects of hydrogen bonding, which are especially important in condensed phases \citep{Frank1957, Elrod1994, Ruckenstein2007}. The ``cooperative'' nature alludes to the fact that the formation of one hydrogen bond promotes the formation of several others and also stabilizes the other bonds within the network. Thus, disrupting these interactions requires breaking the entire network, requiring more energy than a non-hydrogen-bonded network \citep{Ruckenstein2007}. The ability of the –OH functional group in MeOH to form hydrogen bonds and participate in the overall H-bonding cooperativity likely also contributes to the overall increase in binding energy and also explains why even in the multi-layer case MeOH has a higher $E_\text{b}$ than MeSH \citep{Dawes2016}.} Confirming {which effect dominates} would require additional calculations {and experiments} that are beyond the scope of this paper, but as discussed below, this is also consistent with our findings from entrapment experiments.

Our sub-monolayer experiments on amorphous water are quite different from previous experiments using gold as the surface instead. Experimentally, sub-monolayer experiments of  $^{12}$CH$_3$SH on a gold substrate conducted by \citet{Liu2002} showed an 80 K difference in desorption peak temperatures, with thinner coverages exhibiting a shift in T$_\text{peak}$ from 120\,K to 200\,K. This further indicates that the amorphous water surface topology likely plays a crucial role in the binding mechanism for MeSH in astrophysically realistic ices. Furthermore, a study used values from \citet{Liu2002} to constrain their computational method that utilized the TST approximation to determine desorption parameters, predicting  a $\nu_\text{TST}$ of  \mbox{1.3 $\times$10$^{18}$ s$^{-1}$} and  $E_{b,\,\text{TST}}$ of 6522\,K, which are $\sim$\,2.5$\times$ more and $\sim$\,2000\,K higher than our experimentally-derived values, respectively \citep{Ligterink2023}. These differences highlight the importance of laboratory-{based studies of} binding energies on a water surface.

\subsection{MeSH vs. MeOH Entrapment Efficiencies}\label{sec:comp_entrap}

In general, we find that MeOH co-desorbs with water much more effectively than MeSH, consistent with MeOH being bonded more strongly to water. However, surprisingly, there is only a negligible amount of MeSH escaping prior to water ice restructuring, while more than 20\% of MeOH escapes in the fiducial experiment. Despite MeSH exhibiting weak binding to water, it is consistently 100\% entrapped.  However, the entrapped MeSH comes off almost entirely during volcano desorption, suggesting that during ice restructuring, {cracks begin to form} within the water matrix  {that are large} enough {for} MeSH to quickly escape the matrix. {This is also consistent with the behavior of MeSH in MeSH:MeOH:H$_2$O experiments, where MeSH slightly precedes the volcano peak in ices with MeOH, likely due to the impact of MeOH on water crystallization\,kinetics.}

{We also find that CO$_2$ and MeOH desorption kinetics are affected by the presence of MeSH; instead of primarily desorbing at the pure volatile peaks or co-desorbing with water, CO$_2$ and MeOH come off mostly at the volcano desorption peak in mixtures with MeSH.  In other words, {it appears that} even small amounts of MeSH can effectively prevent other matrix constituents from desorbing until the onset of MeSH escaping during volcano desorption. This effect persists even in mixtures of 1:4:35 and 1:8:33 (MeSH:MeOH:H$_2$O), which are comparable to the MeSH:MeOH ratio in comet 67P/C–G of 1:5.5 \citep{Calmonte2016, Schuhmann2019}.} {However, a more detailed experimental follow-up is needed to clarify why and how MeSH influences the desorption of other  matrix components and to evaluate the robustness of this mechanism.}

Together, these experiments suggest that for MeSH, molecular size plays a more significant role in entrapment relative to its binding energy, which is somewhat surprising considering the difference in size between MeSH and MeOH is only around $\sim$15–20\% (see Appendix \ref{app:g16} for relative size estimation). As a result, we can place an upper limit on the typical pore size of compact amorphous water, estimating it to have a diameter smaller than the size of MeSH ($\sim$3.3\,\AA). Overall, the constraints from the desorption and entrapment experiments provide a consistent picture, where ice topology plays a major role for larger molecules such as MeSH, suggesting that we could potentially expect similar behavior from other larger organics and/or S-species that do not participate in {strong} hydrogen bonding.

\subsection{Astrophysical Implications}\label{sec:snowlines}

In this subsection we use our experimentally derived desorption and entrapment characteristics of MeXH species to derive their snow line locations in a fiducial protoplanetary disk given different assumptions about the local icy grain composition. 

We use the disk model from \citet{Oberg2019}, which assumes a disk environment similar to the Solar Nebula. The resulting midplane temperature and density power law profiles normalized to 1 or 2 au are:

\begin{equation}\label{eqn:Tmid}
T_\text{mid}(r) = 140\,\text{K}\bigg[\frac{r}{2~\text{au}}\bigg]^{-0.65},
\end{equation}

and

\begin{equation}\label{eqn:SigmaH}
\Sigma_\text{H}(r) =1500\,\text{g cm$^{-2}$}\bigg[\frac{r}{1\text{\: au}}\bigg]^{-1.5},
\end{equation}


where $r$ is the disk radius in astronomical units (au). We then use the prescription from \citet{Hollenbach2009} to calculate the freeze-out temperature ($T_{\text{f},\,i}$) for a particular species $i$, where each combination of molecule (MeSH or MeOH) and ice environment (pure, layered on H$_2$O, loosely entrapped in H$_2$O, and co-desorbing with H$_2$O) is a different species. By setting the molecular rates of adsorption and desorption on a grain surface equal, we get 

\begin{equation} \label{eqn:Tdes_hollenbach}
T_{\text{f},\,i}(r) \simeq E_{b,\,i}\,\ln\bigg[\frac{4\,N_{i}\,f_{i}\,\nu_{i}}{n_{i}\,v_{\text{th},\,i}(r)}\bigg]^{-1},
\end{equation}

\begin{figure}
\centering
\includegraphics[width=\columnwidth ]{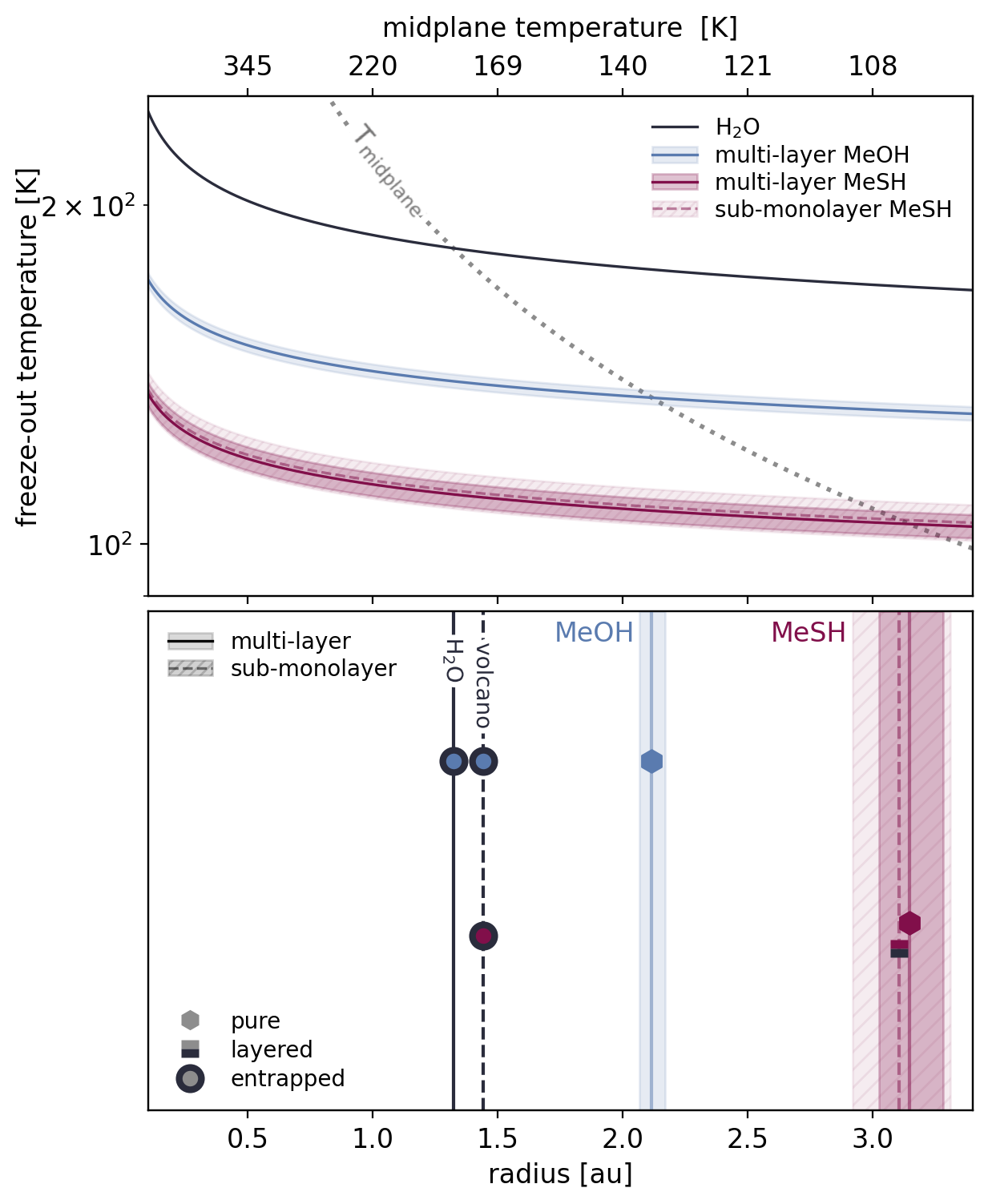}\\
\caption {\textit{Top:} Freeze-out temperatures for molecules as a function of radius/midplane temperature for H$_2$O, MeOH, and MeSH. The assumed midplane temperature profile is overplotted, and the point at which the midplane and freeze-out temperatures are equal is where the snow line of species $i$ is.  \textit{Bottom:} Cartoon illustrating the location of different MeXH snow lines with markers indicating the types of ices studied in this work. } \label{fig:snowlinepreds}
\end{figure}

where $E_{b,\,i}$ is the binding energy of species $i$, $N_{i} $ is the number of adsorption sites per cm$^{2}$ (fixed to $10^{15}$), $f_{i}$ is the fraction of the adsorption sites occupied by species $i$, $\nu_{i}$ is the attempt frequency of the species $i$ in s$^{-1}$, $n_i$ is the gas-phase number density of species $i$, and $v_{\text{th},\,i}$ is the thermal speed of species $i$.

Since both MeSH and MeOH abundances in T Tauri disks are unknown, to estimate $f_{i}$, we use cometary abundances with respect to H$_2$O. In comet 67P/C–G, the abundances with respect to water are: 
$^{12}$CH$_3$SH/H$_2$O\,=\,3.8$\times10^{-4}$  and  $^{12}$CH$_3$OH/H$_2$O\,=\,2.1$\times10^{-3}$ \citep{Calmonte2016, Schuhmann2019}.  We assume H$_2$O/H to be 1.6$\times10^{-4}$. For H$_2$O, we use values of $\nu$ = 4$\times10^{13}$ s$^{-1}$ and 5800 K \citep{Fraser2001} and for the MeXH species, we use the recommended values  in Table \ref{tab:rec_BEs}. To estimate the volcano (i.e. loose entrapment in H$_2$O) snow line location with respect to the water snow line, we assume that the $\sim$\,10 K relative difference between the volcano and H$_2$O co-desorption temperatures can be applied to midplane temperatures. 

Using the assumptions and equations above, we plot the freeze-out temperature as a function of radius/midplane temperature for different species and the resulting snow lines in Figure \ref{fig:snowlinepreds}. The spread (shaded regions) in freeze-out temperatures are due to the $E_{b,\,\text{TST}}$ errors. In the bottom panel of Figure \ref{fig:snowlinepreds}, locations of the MeXH snow lines are shown as a function of radius/midplane temperature, along with markers denoting what kind of ice type would be sublimating at the different locations.

For MeSH, the snow line locations depend completely on whether MeSH is mainly embedded in the water ice phase or resides in a separate ice phase. If embedded, MeSH will desorb near  the water snow line at $\sim$\,173 K; if separate, it desorbs at around 105 K, about twice as far out. MeOH in mixed ices would sublimate at a combination of the volcano and H$_2$O co-desorption  ($\sim$\,173\,–183 K) snow lines, while MeOH in a separate phase, unmixed with H$_2$O,  would desorb at $\sim$\,135 K, interior to the MeSH snow line. Although we did not plot the layered MeOH values since we were unable to derive the MeOH$-$H$_2$O binding energy, qualitatively, we would expect MeOH desorbing off of water grains to coincide with the H$_2$O and volcano snow lines. {If MeOH and MeSH are present in a matrix together. we would expect the MeOH snow line to be pushed towards the H$_2$O  volcano snow line location.} Which snow line location is more accurate depends on the main formation pathway of MeSH (and MeOH) in molecular clouds, since this will determine whether they reside in a CO-rich or H$_2$O-rich ice; for example, MeOH can in either form from H addition to CO ice or O insertion/photochemistry in water-rich ices \citep{Fuchs2009, Bergner2017, Wada2006, Carder2021}.

Finally, the pure and mixed ice spectra show that MeSH displays distinct IR bands (see Appendix \ref{app:spectra}), which shift by up to \mbox{$\sim$0.03 $\mu$m}, dependent on whether MeSH is present in a water-rich or pure ice phase. Such a shift can be easily resolved by the James Webb Space Telescope's (JWST) NIRSpec instrument at $\lambda\sim\,$4.0$\,\mu$m \citep[based on its resolving power of 2700;][]{Jakobsen2022}; inferring mixed ice compositions by comparing JWST NIRSpec and experimental data was recently demonstrated in \citet{Bergner2024}.

\section{Conclusions}
\label{sec:conclusions}

We present a series of experiments to characterize the thermal desorption kinetics and entrapment behavior of methyl mercaptan (MeSH), the simplest complex organosulfur, ices for the first time. We also contextualize all results with its O-bearing analog, methanol (MeOH). These results provide fundamental astrochemical model inputs and reveal some peculiarities of the organosulfur desorption and entrapment kinetics. In summary, we provide the first experimental desorption constraints for MeSH by analyzing three different ice types (pure, layered, and mixed) to obtain binding energies, attempt frequencies, and entrapment efficiencies, and our main results are as follows:

\begin{enumerate}
    \item We find the transition state theory (TST) model is the best approximation for constraining attempt frequencies, which are necessary to derive binding\,energies. 
    \item The derived multi-layer MeSH$-$MeSH and MeOH$-$MeOH binding energies are 4610 $\pm$ 110 K and 5750 $\pm$ 80 K, respectively. The derived sub-monolayer MeSH$-$H$_2$O binding energy, 4640 $\pm$ 170 K,  is remarkably similar to the multi-layer indicating that MeSH desorbs at the same temperature regardless of whether it is in a pure or water matrix, highlighting its distinct behavior compared to MeOH.
    \item Most notably, we find that even though MeSH does not bind well to water, it is nearly 100\% entrapped in mixed water-dominated ice matrices regardless of thickness and composition, and it comes off almost exclusively at the volcano desorption peak.  
    \item {The presence of MeSH inhibits the desorption of both CO$_2$ and MeOH by increasing their entrapment (up to 76\% and \,96–100\% in the cases of CO$_2$ and MeOH, respectively) within the water matrix, with both following desorption with MeSH during water crystallization.}
    \item {We show, for the first time, how a molecule's size significantly affects its own entrapment efficiency and influences the entrapment and retention of smaller molecules in H$_2$O-dominated mixtures.}  
    \item {These findings imply that the difference in size between MeSH and MeOH—which is only on the order of $\sim$15–20\% and $\sim$0.8\,\AA—is enough to inhibit the diffusion of MeSH through pores in the water matrix, allowing us to place an upper limit on compact water's pore size of 3.3\,\AA.}
    \item In Solar-like midplane conditions, MeSH sublimation occurs at midplane temperatures of 105 K, but may also co-exist with water up until 173 K, dependent on whether it formed mixed with water or in a separate phase.
\end{enumerate}

\vspace{3mm}
{We thank the reviewers for their insightful recommendations and feedback.} S.N. would like to thank Julia C. Santos and Jennifer B. Bergner for support with  sub-monolayer modeling, and Reggie Hudson for sending us the amorphous $^{12}$CH$_3$SH spectrum for our comparison studies.  Funding support is acknowledged from NSF GRFP grant No. 2236415 (S.N.), the P.E.O. Scholar Award (S.N.), a grant from the Simons
Foundation (686302, K.I.\"O.), and an award from the Simons Foundation (321183FY19, K.I.\"O.).

\software{{\fontfamily{lmtt}\selectfont{Matplotlib}}  \citep{Hunter2007},\, {\fontfamily{lmtt}\selectfont{NumPy}} \citep{VanDerWalt2011},\, {\fontfamily{lmtt}\selectfont{SciPy}} \citep{2020SciPy-NMeth},\, {\fontfamily{lmtt}\selectfont{Pandas}} \citep{pandas},\, {\fontfamily{lmtt}\selectfont{lmfit}}  \citep{lmfit},\, {\fontfamily{lmtt}\selectfont{statsmodels}} \citep{statsmodels},\, and {\fontfamily{lmtt}\selectfont{Gaussian 16}} \citep{g16}.}

\bibliography{main}
\bibliographystyle{aasjournal}

\appendix
\restartappendixnumbering
\vspace{-5mm}

\section{Computational Methods: Gaussian 16 Calculations}\label{app:g16}

\label{sec:computational}

To supplement our analysis and interpretation of the experimental results, we perform complementary ab initio electronic structure calculations with optimization of the geometry and frequency using Gaussian 16 (G16) \citep{g16}. We use the results to calculate essential parameters needed to derive binding energies using  the transition state theory (TST) model (see \S\ref{sec:nu_TST} for details){,} to determine dimer binding energies {which} aid in interpretation of the experimental results{, and to quantify isotope effects on the optimized molecular geometry, dimer binding energies, and relative band strengths}. Previous work \citep[e.g.,][]{Wakelam2017, Das2018, Piacentino2022, Woon2021} that benchmarked the performance of basis sets and cluster types informed our choices of methods within density functional theory (DFT). We performed our calculations at the {\fontfamily{lmtt}\selectfont{M06-2X/aug-cc-pVQZ}} (for obtaining parameters such as principal moments of inertia and bond lengths),  {\fontfamily{lmtt}\selectfont{M06-2X/aug-cc-pVDZ}} (for estimating the binding energies of dimers){, and {\fontfamily{lmtt}\selectfont{B3LYP/aug-cc-pVDZ}} (for determining key vibrational modes and band strengths)} levels of theory \citep{Dunning1989, Woon1993, Becke1993, Lee1988}.

\subsection{{Calculation of Optimized Molecular Geometries}}\label{app:optstructure}

\begin{figure*}[ht!]
    \centering
     \includegraphics[width =0.75\textwidth]{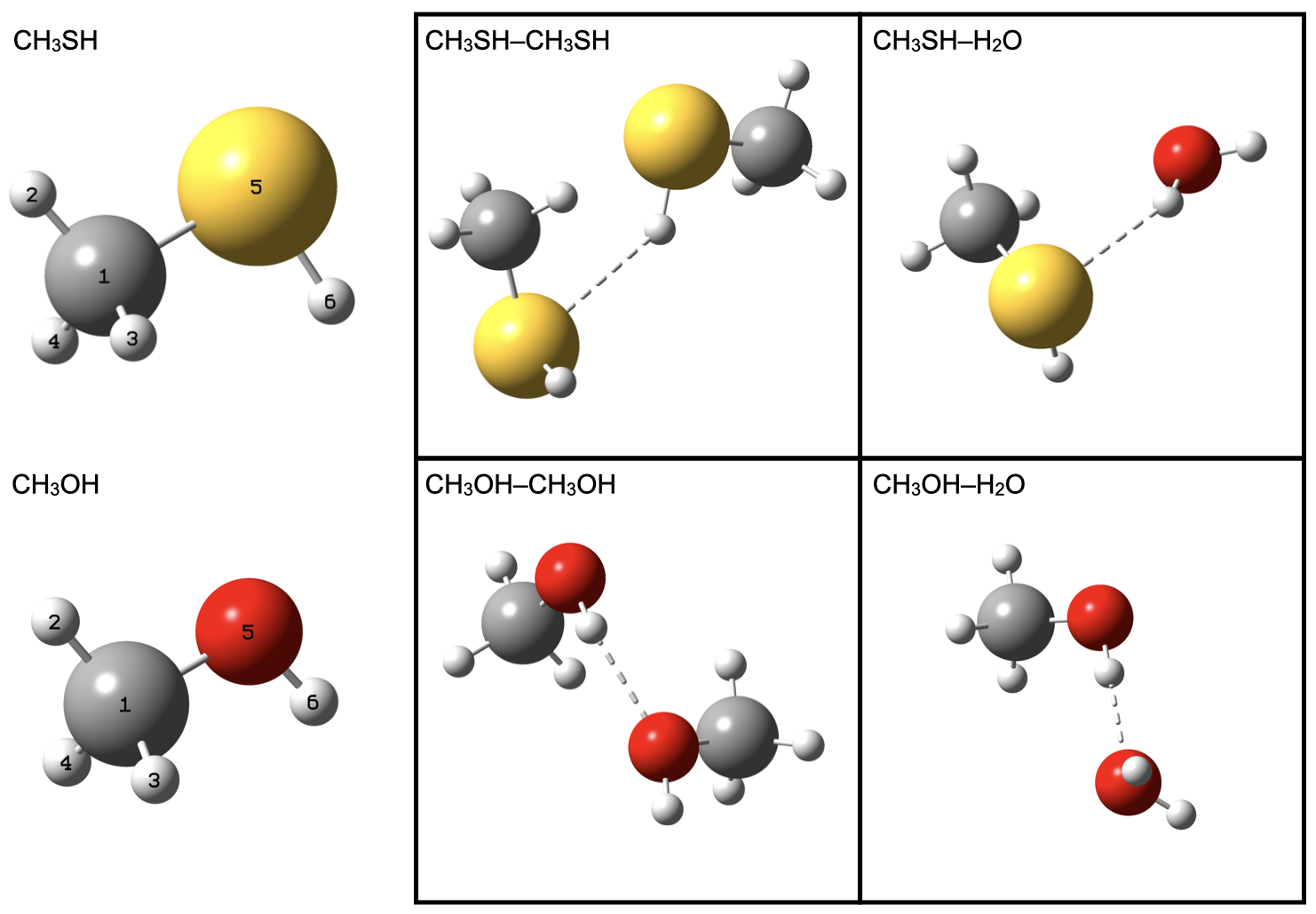}
    \caption{{The left-most numbered molecules are the optimized molecular geometries for CH$_3$SH (top, yellow) and CH$_3$OH (bottom, red)}, scaled to size based on covalent interactions. Atoms are numbered for clarity and are used in reference to relevant {atomic pair distances} listed in Table {\ref{tab:computational_properties}}. {In the boxes are the l}owest-energy dimer geometries for CH$_3$SH (top) and CH$_3$OH (bottom), scaled to size based on covalent interactions. The closest bonding interaction is depicted as a dashed line. Note the differences in orientation between respective CH$_3$XH dimers. {The different orientations and energies were the same for both $^{12}$C- and $^{13}$CH$_3$XH dimers.}} \label{fig:MeXH_g16comparison}
\end{figure*}

While precise collisional cross-sections are unavailable, we can estimate the relative size difference using computationally optimized molecular geometries. To estimate the pore size (see \S\ref{sec:comp_entrap}), we use the largest length of the MeXH molecule, which spans from the H of the thiol/alcohol functional group to the H on the methyl group that is furthest away (labeled as atoms 2 and 6 in Figure \ref{fig:MeXH_g16comparison}). We performed these calculations {at the {\fontfamily{lmtt}\selectfont{M06-2X/aug-cc-pVQZ} level of theory}} for both the $^{12}$C and  $^{13}$C isotopologues and found negligible differences. All relevant bond lengths are presented in  Table \ref{tab:computational_properties} to highlight the size differences between the two molecules. This bond length is 3.29\,\AA\, for MeSH vs. 2.82\,\AA\, for MeOH which results in a $\sim$15–20\% difference in size. {We also used these calculations to determine the principal moments of inertia ($I_x$, $I_y$, and $I_z$) and symmetry factor ($\sigma$), which a}s explained in \S \ref{sec:nu_TST}, are required to approximate the attempt frequency, $\nu$, using the TST model. The values used to determine the recommended $\nu_\text{TST}$ in Table \ref{tab:rec_BEs} {take in the $^{12}$CH$_3$XH values} listed in Table \ref{tab:computational_properties}. {We calculated $\nu_\text{TST}$ using both the $^{12}$C and $^{13}$C-H$_3$XH moments of inertia and found $<$\,5\% difference, making the $T_\text{peak}$ value as the primary source of uncertainty in $\nu_\text{TST}$. }

\begin{deluxetable}{lllccccc}
\tablecaption{{Summary of all computational calculations and resulting properties used in this work. The bond distances refer to the numbered atoms shown in Figure \ref{fig:MeXH_g16comparison}. The principal moments of inertia and symmetry factors are used in the $E_{b,\,\text{TST}}$ calculations. The computationally-derived binding energies ($ E_{b,\,\text{comp}}$) are calculated using Eq. \ref{eqn:dimer}. The band strengths are shown only for the vibrational modes which are used to calculate the ice column densities (see Table \ref{tab:bandstrengths}) or are most affected by the $^{13}$C isotope.}}
\label{tab:computational_properties}
\tablehead{\colhead{Calculation Type} & \multicolumn{1}{c}{Property}  & \colhead{Parameter} & \multicolumn{2}{c}{{CH$_3$SH (X\,=\,S)}} & \multicolumn{2}{c}{{CH$_3$OH (X\,=\,O)}} &  \colhead{Level of Theory} \\ 
\cmidrule(l{2pt}r{2pt}){4-5} \cmidrule(l{2pt}r{2pt}){6-7} \nocolhead{}  & \multicolumn{1}{c}{} & \nocolhead{} &   \multicolumn{1}{c}{$^{12}$C} & \multicolumn{1}{c}{$^{13}$C} & \multicolumn{1}{c}{$^{12}$C} & \multicolumn{1}{c}{$^{13}$C} & \nocolhead{} }
\startdata
\multirow{9}{*}{Optimized Geometry}  & \multirow{5}{*}{\parbox{2.78cm}{Atomic\,Pair Distance [\AA]}}   & X[5]–C[1] & 1.81 & 1.81 & 1.41 & 1.41 & 
\multirow{9}{*}{{\fontfamily{lmtt}\selectfont{M06-2X/aug-cc-pVQZ}}} \\ 
& & X[5]–H[6] & 1.34 & 1.34 & 0.96 & 0.96 &   \\ 
& & C[1]–H[2,3,4] & 1.09 & 1.09 & 1.09 & 1.09 &   \\ 
& & H[2]–H[6] & 3.29 & 3.29 & 2.82 & 2.82 &  \\ 
& & H[3,4]–H[6] & 2.67 & 2.67 & 2.35 & 2.35 &  \\ 
\cmidrule{2-7}
& \multirow{3}{*}{\parbox{2.6cm}{Moments of Inertia [amu \AA$^2$]}} & $I_x$ & 4.87 & 4.87 & 3.91 & 3.91 & \\ 
&  & $I_y$ & 38.87 & 40.34 & 20.21 & 20.72 &  \\ 
&  & $I_z$ & 40.53 &  42.01 & 20.94 & 21.45 & \\ 
\cmidrule{2-7}
& \multirow{1}{*}{Symmetry Factor}  & $\sigma$ & 1 & 1 & 1 & 1 & \\
\hline
\hline
\multirow{2}{*}{Dimer Binding Energy}  & \multirow{2}{*}{$ E_{b,\,\text{comp}}$ [K]} & CH$_3$XH–CH$_3$XH  & 1642 & 1642 & 3105 & 3105 &
\multirow{2}{*}{{\fontfamily{lmtt}\selectfont{M06-2X/aug-cc-pVDZ}}}\\ 
&  & CH$_3$XH–H$_2$O & 2588 & 2588 & 3033 & 3033 &  \\ 
\hline
\hline
\multirow{6}{*}{\parbox{2.4cm}{Band\,Strength [cm molecule$^{-1}$]}} & \multirow{2}{*}{S–H stretch\,/\,$\times$10$^{-19}$} & harmonic & 6.36 & 6.35 & – & – & \multirow{6}{*}{{\fontfamily{lmtt}\selectfont{B3LYP/aug-cc-pVDZ}}} \\
 & & anharmonic & 7.08 & 7.07& – & – &  \\ \cmidrule{2-7}
 & \multirow{2}{*}{C–S stretch\,/\,$\times$10$^{-19}$} & harmonic & 3.93 & 3.75 & – & – & \\
 & & anharmonic & 4.19 & 4.02& – & – &  \\
 \cmidrule{2-7}
 & \multirow{2}{*}{C–O stretch\,/\,$\times$10$^{-17}$} & harmonic & – & –  & 1.92 & 1.76 \\
 & & anharmonic & – & – & 1.98 & 1.78 &\\
 \hline
\hline 
 \enddata
\end{deluxetable}

\FloatBarrier

\subsection{Calculation of Dimer Binding Energies}\label{app:dimer}

As shown in \citet{Piacentino2022}, in many cases, dimer calculations to extract binding energies are able to reproduce experiments well. Since we are only using these calculations as a reference to understand the experimental results, we do not model the binding energy using larger water or MeXH clusters. All binding energies are calculated using {\fontfamily{lmtt}\selectfont{M06-2X/aug-cc-pVDZ}} level of theory. To extract binding energies computationally ($E_{b,\,\text{comp}}$) between a molecule A and molecule B, we use the following equation:  

\begin{equation}\label{eqn:dimer}
    E_{b,\,\text{comp}} = E_{A+B} - (E_A+ E_B), 
\end{equation}

where $E_{A+B}$ is the energy of the dimer and $E_{i}$ is the energy of the molecule $i$. For the dimer optimization, we explored different initial configurations and chose the lowest energy of the optimized dimer geometry to calculate the binding energy. Following the methods described by \citet{Wakelam2017} and \citet{Piacentino2022}, we have not included the zero-point energy correction. We report the computational binding energies in Table \ref{tab:computational_properties} {and note no differences between the respective $^{12}$C and $^{13}$C-H$_3$XH computationally-determined binding energies}.

\FloatBarrier

\subsection{{Calculation of Band Strengths}}\label{app:bandstrengthcalc}

{In order to further validate the assumption that the $^{12}$CH$_3$XH band strengths could be used for our $^{13}$CH$_3$XH experiments, we calculated the key vibrational modes and band strength intensities—namely the C–X and S–H stretches—which are either used to calculate ice column densities and/or are most affected by the carbon isotope. We performed these calculations at both the {\fontfamily{lmtt}\selectfont{M06-2X/aug-cc-pVDZ}}  and 
{\fontfamily{lmtt}\selectfont{B3LYP/aug-cc-pVDZ}} levels of theory and found the latter to be better at determining key vibrational modes, consistent with literature \citep[see e.g.,][]{Woon2021}. We include both the harmonic and anharmonic results to emphasize that, in both cases, the relative variation in band strengths between the isotopologues is at most 10\% (which occurs for the case of $^{12}$C vs. $^{13}$CH$_3$OH). Of note here is that the variation in band strength intensities for the $^{12}$C vs. $^{13}$CH$_3$SH isotopologues is  0.06\% and 0.16\% for the S–H stretch, and 4.54\% and 3.95\% for the  C–S stretch, in the harmonic and anharmonic cases, respectively. It makes sense that the C–S stretch variation is larger, as it is more directly impacted by the isotope. All of these differences are smaller than the variation calculated for the C–O stretch in the CH$_3$OH case, where we find a difference of 8.37\%	and 10.05\% in the harmonic and anharmonic cases, respectively. These results clearly demonstrate that using the S–H stretch band strength from the literature to calculate ice column densities is appropriate and that a 20\% assumed band strength error due to applying $^{12}$C band strengths to $^{13}$CH$_3$XH data is a conservative estimate.} 

\section{IR Spectra}\label{app:spectra}

\subsection{Pure $^{12}$C- vs. $^{13}$CH$_3$SH Spectra}

To ensure we can use the $^{12}$C-methyl mercaptan band strength for the $^{13}$C isotopologue,  we overplot the pure $^{12}$CH$_3$SH \citep[from][]{Hudson2016} and $^{13}$CH$_3$SH, or MeSH, (this work) spectra in {\textit{the top panel of}} Figure \ref{fig:comparingspectra}, with a{inset that zooms in} on the S-H stretch that is used for determining the ice column densities. The shapes of the S-H feature are nearly identical, showing that the S-H stretch is largely unaffected by the C isotope supporting our use of the available $^{12}$CH$_3$SH band strength. Although there are some minor shifts for other peaks, presumably due to the isotope, assigning these features is beyond the scope of this work.

\subsection{Ternary MeSH Spectra}

\begin{figure*}[ht!]
    \centering
     \includegraphics[width = \columnwidth]{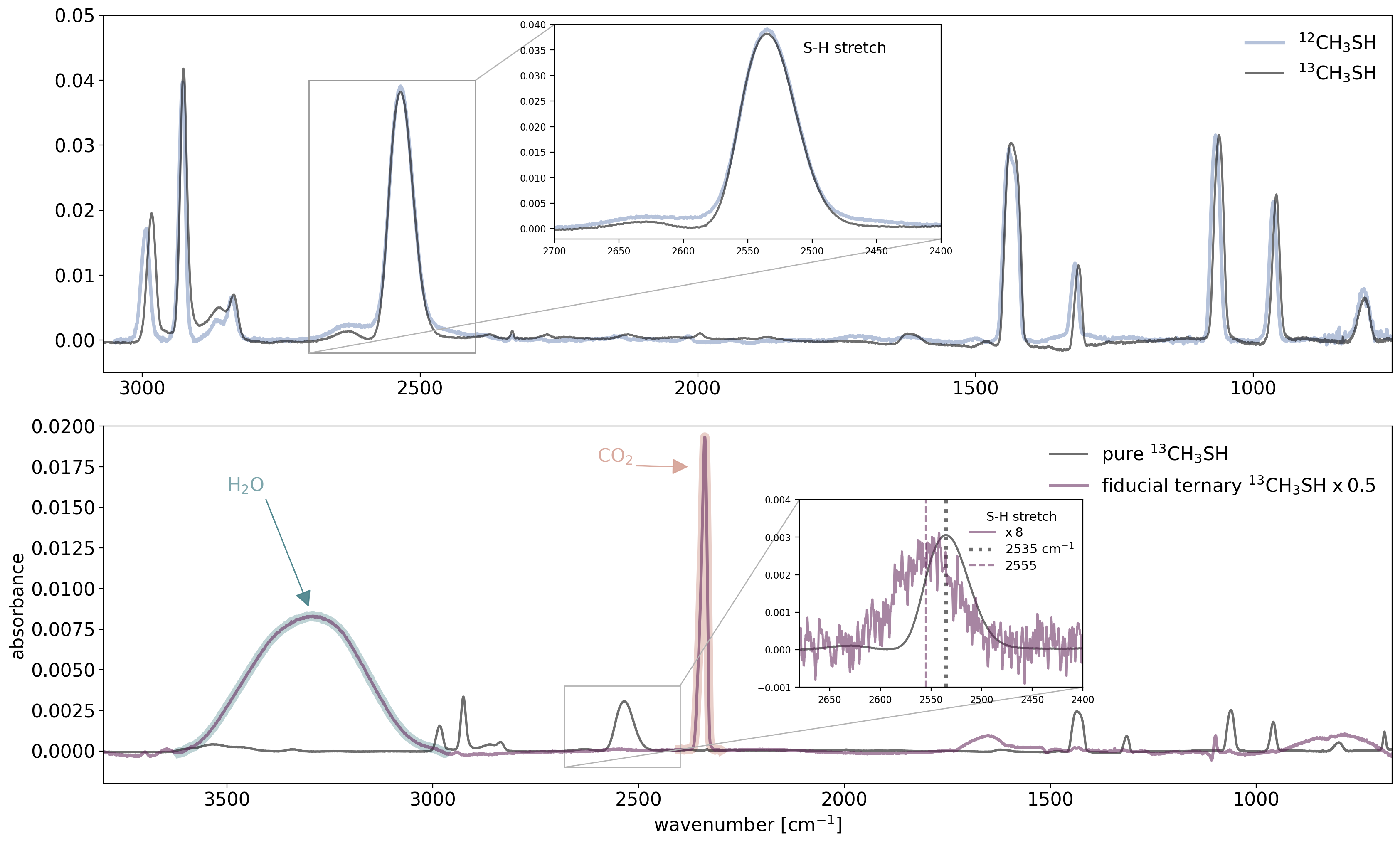}

    \caption{\textit{Top:} Comparison of $^{12}$C- and $^{13}$CH$_3$SH spectra; the $^{12}$CH$_3$SH spectrum is from \cite{Hudson2016} and is reproduced with permission. The main panel shows the overall spectra for both isotopologues, while the inset zooms in on the S–H stretching region that is used for determining ice coverages. The shape of this feature remains unchanged between the two isotopologues. \textit{Bottom:} Comparison of spectra of pure MeSH  (Expt. 5 from Table \ref{tab:thermal_expt_summ}) and the fiducial ternary MeSH (Expt. 2{0} from Table \ref{tab:entrap_expt_summ}) showing the $\sim$\,20 cm$^{-1}$ (or $\sim$\,0.03 $\mu$m)  shift for the S–H stretching feature that is used to determine ice coverages. The H$_2$O and CO$_2$ features used to determine the ice coverages for all entrapment experiments are also highlighted in light blue and orange, respectively. The main panel shows the {full wavenumber range} for both spectra, while the inset zooms in on the S–H stretching region highlighting the shift. The ternary feature shows both a broadening as well as a shift in the peak position.} \label{fig:comparingspectra}
\end{figure*}

 In {the bottom panel of} Figure \ref{fig:comparingspectra}, we show that the pure and mixed ternary MeSH ices display distinct IR bands that correspond to a 0.03\,$\mu$m shift. 

\FloatBarrier

\subsection{Multi-Layer Baseline Corrections and Fits}

\begin{figure*}[ht!]
    \centering
     \includegraphics[width =0.85\columnwidth]{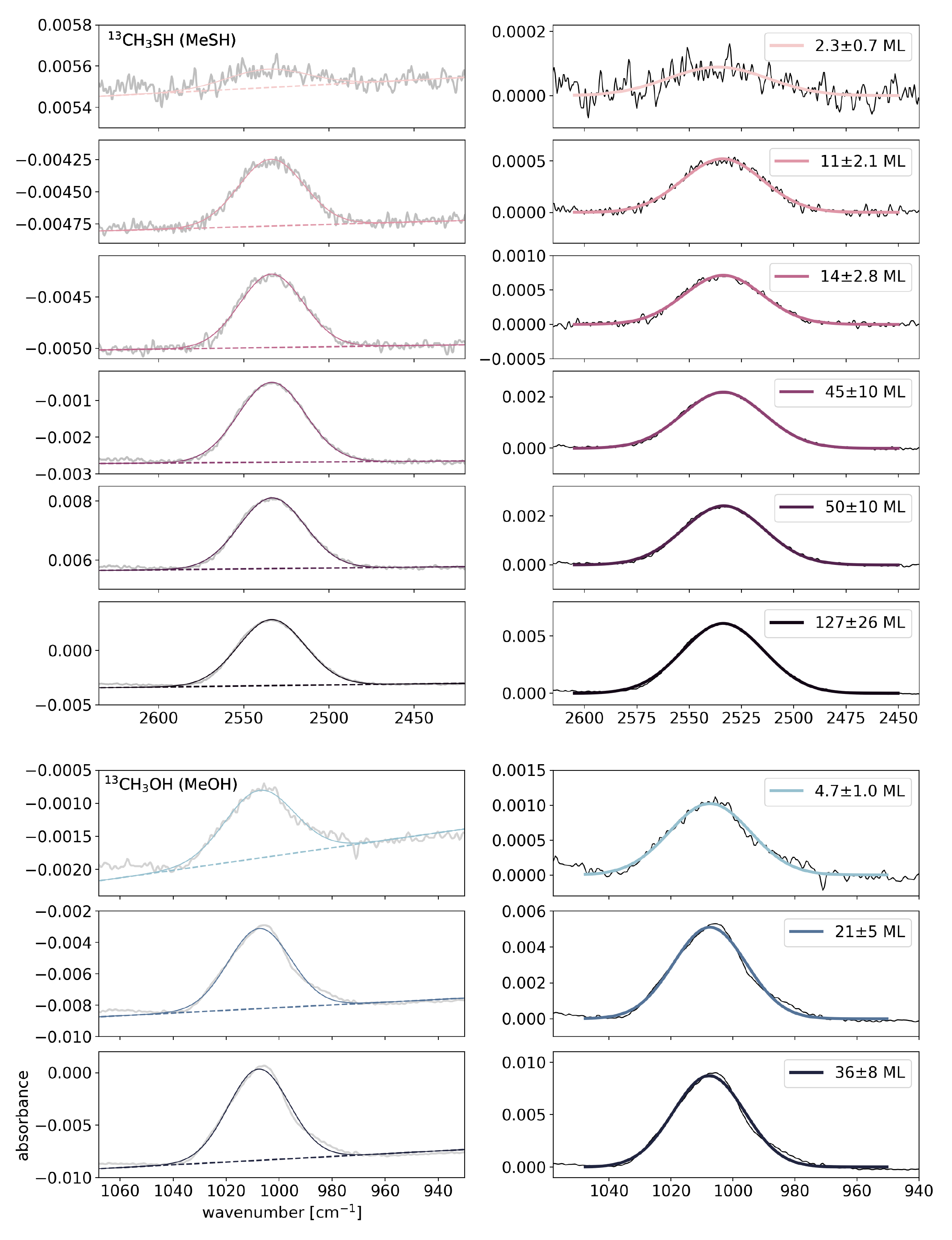}
    \caption{Multi-layer Me{X}H IR data with {MeSH in shades of pink (top 6 rows corresponding to Expts. 1–6 in Table \ref{tab:thermal_expt_summ} from top to bottom) and MeOH in shades of blue (bottom 3 rows corresponding to Expts. 7–9 in Table \ref{tab:thermal_expt_summ} from top to bottom). The panels are} zoomed-in to the stretching mode region used to quantify ice coverage thickness. Given we are approaching the detection limit of the IR feature for the 2.3 ML (top-most) experiment, we assume a 30\% error rather than the 20\% error assumed for all other experiments. \textit{Left columns:} Raw spectra overlaid with the Gaussian and linear baseline fits. \textit{Right columns:} Corrected spectra with corresponding ice coverages and uncertainties indicated.} \label{fig:MeXH_spectra}
\end{figure*}

In Figure \ref{fig:MeXH_spectra}, we show the strongest IR features (used to determine ice coverages for MeXH ices corresponding to the respective stretches listed in Table \ref{tab:bandstrengths}). These spectra are used for extracting multi-layer binding energies and for creating a calibration curve used to determine the sub-monolayer coverages (see Appendix \ref{app:subML_calc}). 

\FloatBarrier


\FloatBarrier

\section{Sub-monolayer Ice Coverage Calculation}\label{app:subML_calc}

\begin{figure*}[ht!]
    \centering
     \includegraphics[width =0.8\columnwidth]{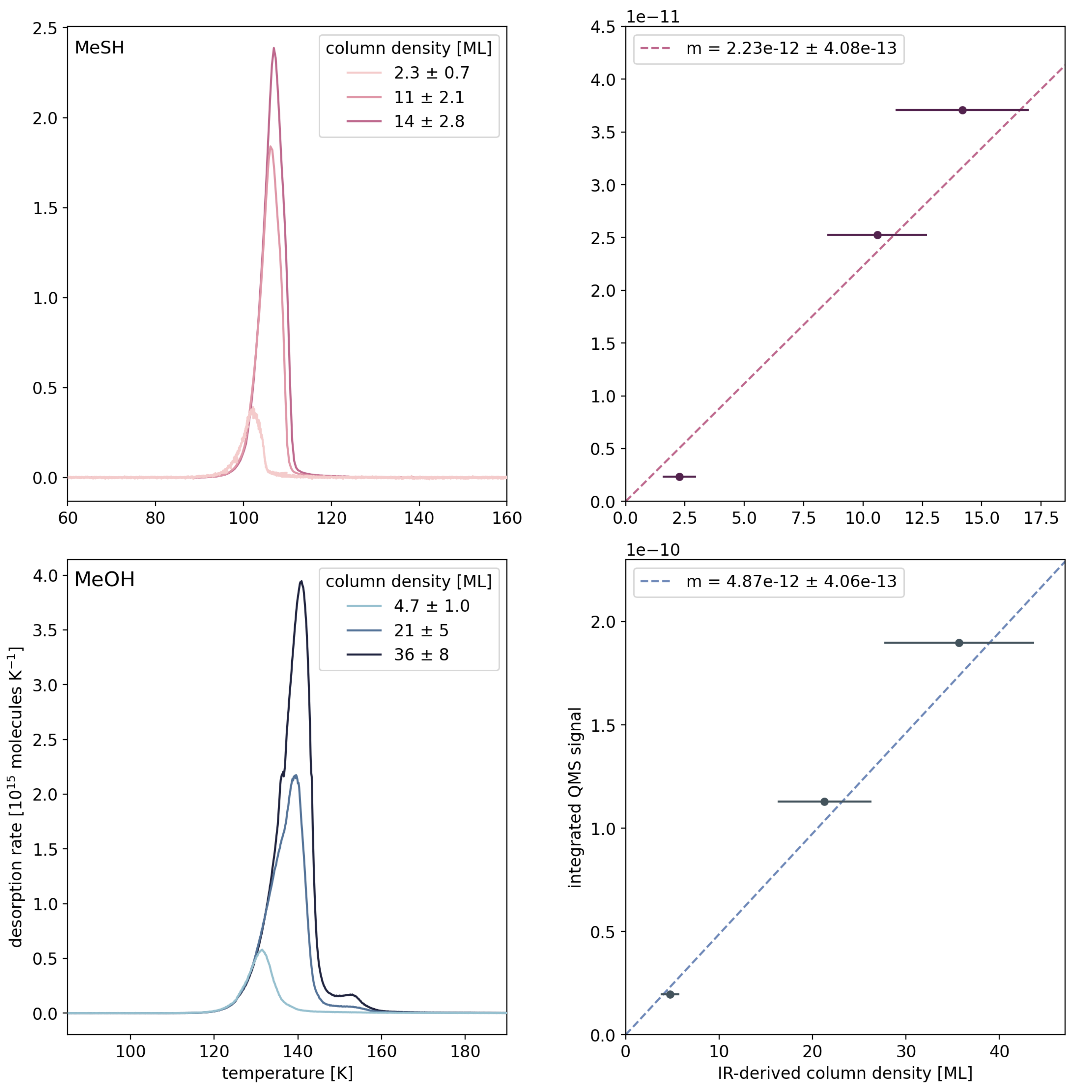}
    \caption{{Pure MeXH TPD curves (left) used to create the calibration curves (right) for MeSH (top row, pink) and MeOH (bottom row, blue). The TPD curves correspond to Expts. 1–3 for MeSH and Expts. 7–9 for MeOH in Table \ref{tab:thermal_expt_summ}. The r}esulting calibration curve shows the integrated QMS signal as a function of IR-derived column density with the best-fit line and derived calibration constant $m$ overplotted.} \label{fig:MeXH_calibrations}
\end{figure*}

As shown in {the top left panel in} Figure \ref{fig:MeXH_spectra}, the thinnest ice ($\sim$\,2 ML) is very noisy and approaches our IR detection limit. Thus, to determine sub-monolayer ice coverages for our layered ices, we created a calibration curve (see right column of Figure \ref{fig:MeXH_calibrations}) to relate the integrated QMS signals  to the IR-derived column densities. To derive a calibration constant, the curve was fit via a weighted least squares (WLS) linear regression algorithm \citep[{{\fontfamily{lmtt}\selectfont{statsmodels.regression.linear\_model.WLS}}} from][]{statsmodels}  where the points were weighted by the typical inverse of the variance squared  ($1/\sigma^2$) where $\sigma$ corresponds to the calculated coverage uncertainties. The resulting calibration constant ($m$) was used to determine sub-monolayer ice coverages as shown in Figure \ref{fig:full_MeSH_subML_tpds} and \ref{fig:full_MeOH_subML_tpds}.  We derive the calibration constant using only the three thinnest MeSH experiments, as these are most relevant for sub-monolayer analyses.

\FloatBarrier

\section{Supplementary $^{13}$CH$_3$OH  Figures}\label{app:MeOH_supp}
All relevant multi-layer and sub-monolayer MeOH plots are shown below; these figures are analogous to those of MeSH in the main text. All key values derived from the analyses that are necessary for discussion are presented in the main text. 

\subsection{MeOH Multi-Layer Experiments}\label{app:MeOH_multilayer}

\begin{figure*}[ht!]
    \centering
     \includegraphics[width =0.52\columnwidth]{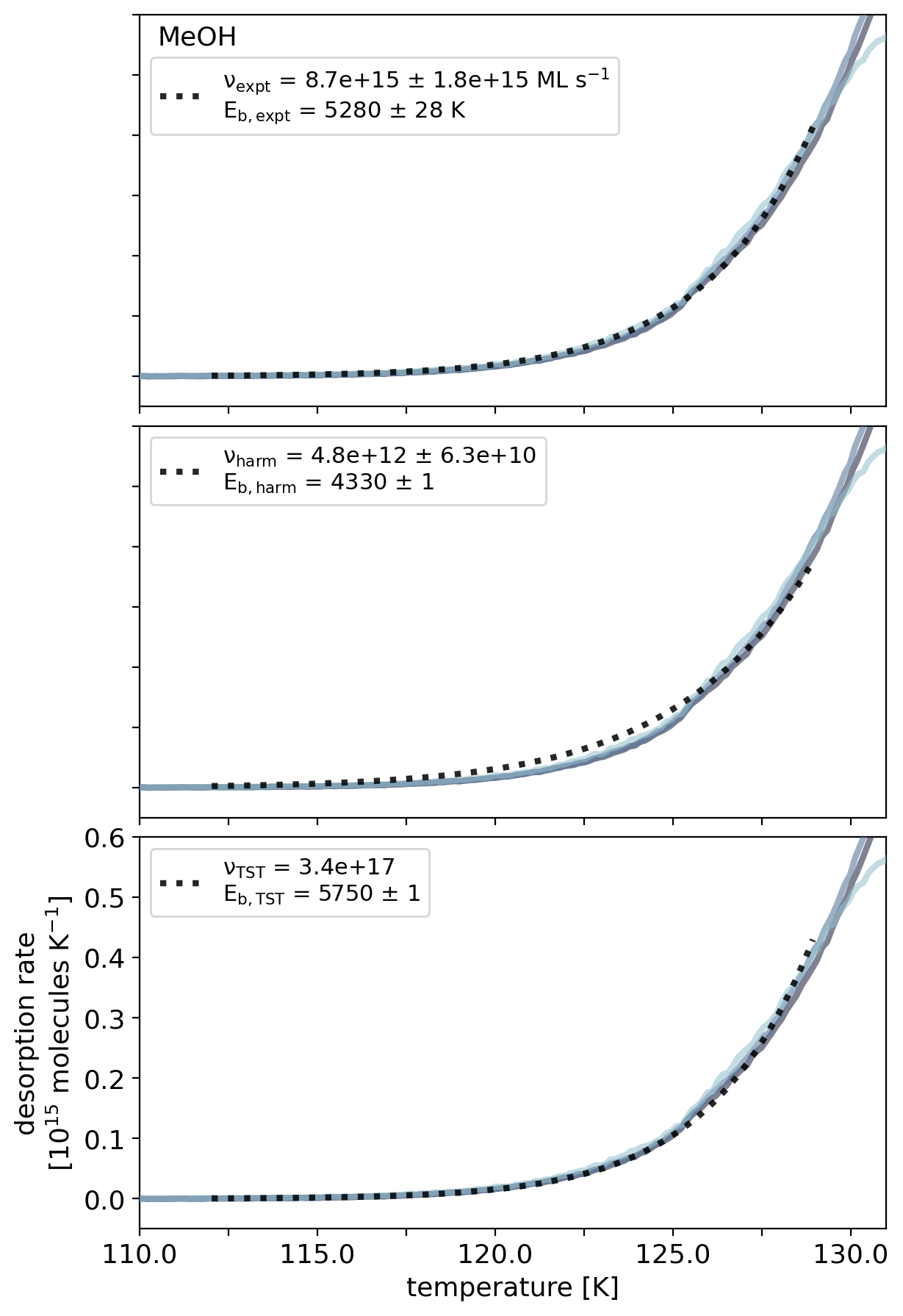}
    \caption{Same as Figure \ref{fig:MeSH_combined_PW_fits} but for MeOH. Visually, the fits for all three methods (described in detail in \S\ref{sec:mult}) are better for MeOH compared to MeSH. However, the harmonic approximation still deviates the most from the leading edges, and the fitting errors on $\nu_\text{expt}$ are 20\%, making the TST approximation the preferred method. The recommended TST-derived values with the associated errors from the fit are presented in Table \ref{tab:rec_BEs}.} \label{fig:MeOH_combined_PW_fits}
\end{figure*}
\FloatBarrier

\begin{figure*}[ht!]
    \centering
     \includegraphics[width =0.7\columnwidth]{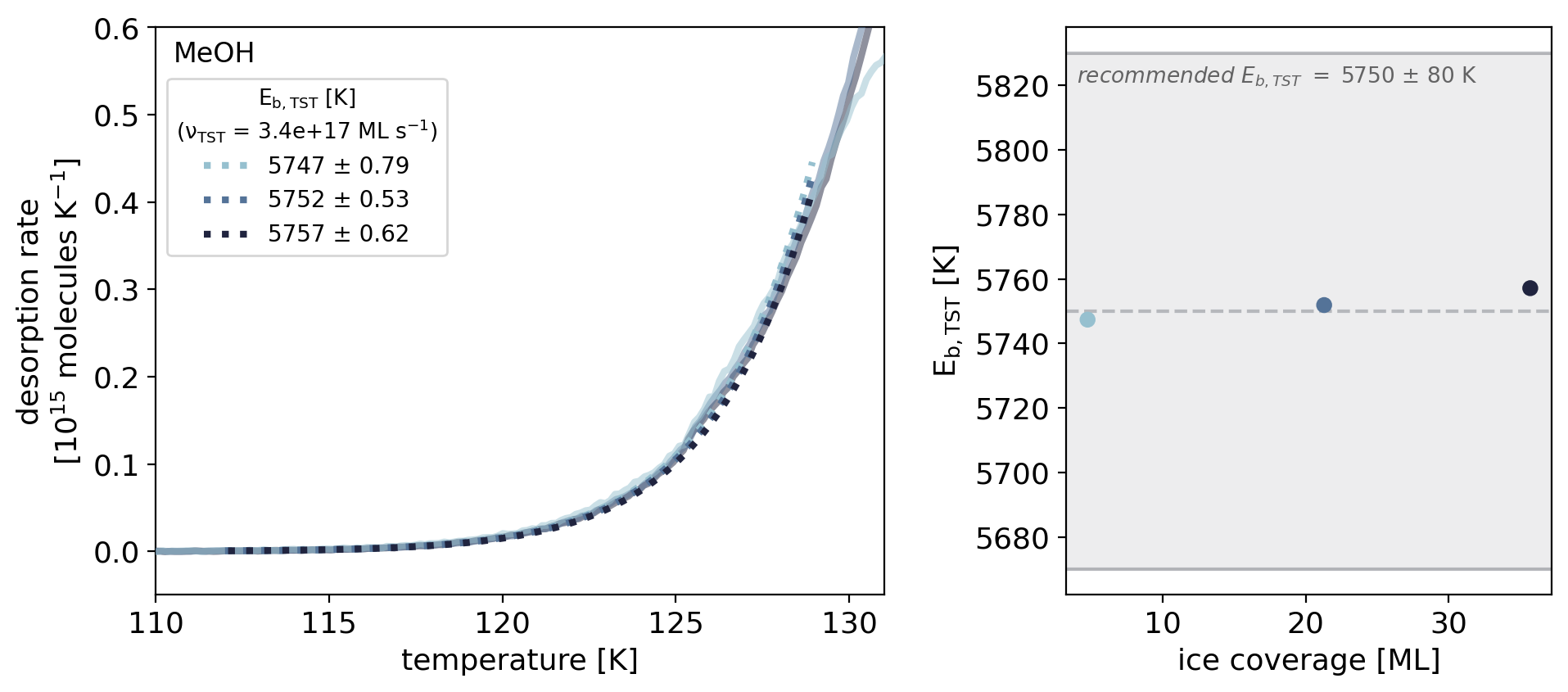}
\caption {Similar to Figure \ref{fig:MeSH_BETSTafoML} but for MeOH (and oriented horizontally). Compared to MeSH, all three individual fits overlap very well, and we find that for a $\sim$\,30 ML difference, the binding energies only deviate by $\sim$10 K and are well within our recommended uncertainties of $\pm$\,80 K.} 
\label{fig:MeOH_BETSTafoML}
\end{figure*}
\FloatBarrier

\subsection{MeOH Sub-Monolayer Experiments}\label{app:MeOH_submonolayer}

\begin{figure*}[ht!]
    \centering
     \includegraphics[width =0.5\columnwidth]{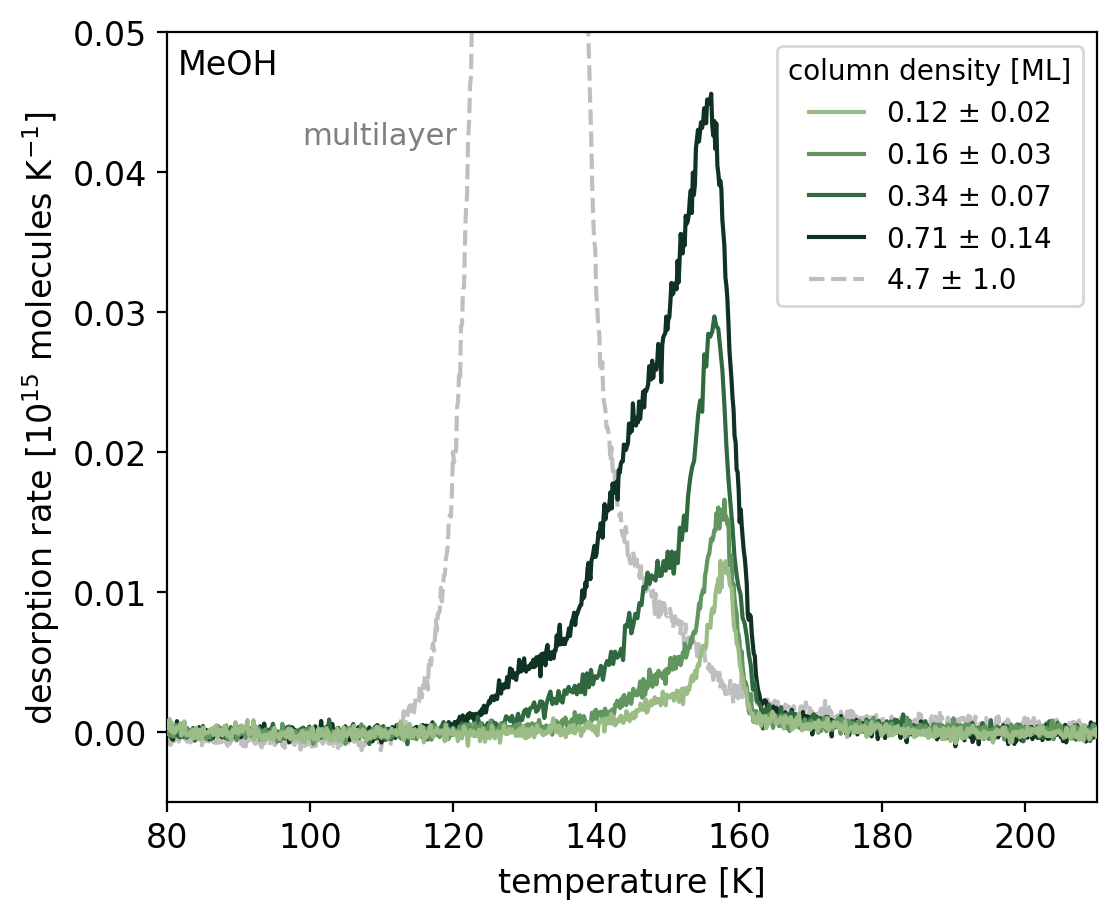}
    \caption{Same as Figure \ref{fig:full_MeSH_subML_tpds}  but for MeOH, corresponding to Expts. 13–16 in Table \ref{tab:thermal_expt_summ}.} \label{fig:full_MeOH_subML_tpds}\end{figure*}

The sub-monolayer TPD curves of MeOH (Figure \ref{fig:full_MeOH_subML_tpds}) show sharp contrast with that of MeSH (Figure \ref{fig:full_MeSH_subML_tpds}). For MeSH we see that the T$_\text{peak}$ remains the same while transitioning from the multi-layer to sub-monolayer regime and the desorption profiles {become} more Gaussian as ices become thinner. However, for MeOH we see that {as}  T$_\text{peak}$ increases, the curves align at the trailing edge (which is distinct from all of the other previous experiments). Qualitatively, this shift suggests that the $E_b$ for MeOH$-$H$_2$O  must be higher than that of MeOH$-$MeOH.  However, because the T$_\text{peak}$ coincides with water co-desorption, it is unclear whether the observed profile reflects the true MeOH$-$H$_2$O bonding or whether it is due to entrapment (see \S\ref{sec:entrapment}). In reality, it could be a combination of these factors, but it is very difficult to disentangle and quantify the contribution of each effect. As a result, we are unable to definitively determine binding energies for layered MeOH$-$H$_2$O.

\FloatBarrier

\FloatBarrier



\end{document}